\documentclass[a4paper,fleqn,usenatbib]{mnras}

\usepackage{newtxtext,newtxmath}

\usepackage[T1]{fontenc}
\usepackage{ae,aecompl}

\usepackage{lineno}
\modulolinenumbers[5]

\usepackage{amsmath}
\usepackage{amssymb}

\usepackage{psfrag} 

\usepackage[utf8]{inputenc} 
\usepackage{graphicx}

\usepackage{relsize}
\usepackage{pbox}
\usepackage[usenames,dvipsnames]{color}
\usepackage{bm}

\usepackage{array}
\usepackage{soul}

\newcommand{\addnew}[1]{{#1}}

\newcommand{\zamo}[1]{#1}
\newcommand{\zze}{0_{n_r\times n_\theta}}
\newcommand{\II}[1]{\mathcal{I}^{#1}}

\newcommand{\OmegaBH}{\Omega_{\rm \text{\tiny BH}}}

\title[Numerically solving GSE in Kerr spacetimes]{Numerically solving the relativistic Grad-Shafranov equation in Kerr spacetimes: Numerical techniques}

\author[J. F. Mahlmann, P. Cerdá-Durán and M. A. Aloy]
{J. F. Mahlmann$^1$\thanks{jens.mahlmann@uv.es}, P. Cerdá-Durán$^1$\thanks{pablo.cerda@uv.es} and M.A. Aloy$^1$\thanks{miguel.a.aloy@uv.es}\\ $^1$Departament d'Astronomia i Astrofísica, Universitat de València, 46100, Burjassot, Spain}

\date{\today. Accepted 3 April 2018. Received 26 March 2018; in original form 2 February 2018}

\pagerange{\pageref{firstpage}--\pageref{lastpage}} \pubyear{2018}

\begin{document}

\maketitle

\label{firstpage}

\begin{abstract}
 The study of the electrodynamics of static, axisymmetric and
 force-free Kerr magnetospheres relies vastly on solutions of the so
 called relativistic Grad-Shafranov equation (GSE). Different
 numerical approaches to the solution of the GSE have been introduced
 in the literature, but none of them has been fully assessed from the
 numerical point of view in terms of efficiency and quality of the
 solutions found. We present a generalization of these algorithms
 and give detailed background on the algorithmic implementation. We
 assess the numerical stability of the implemented algorithms and
 quantify the convergence of the presented methodology for the most
 established setups (split-monopole, paraboloidal, BH-disk, \addnew{uniform}).
\end{abstract} 


\section{Introduction}
\label{sec:introduction}

The so called Grad-Shafranov equation (GSE) \citep{Luest1954,
 Grad1958,Shafranov1966} appears as the master equation to determine
axisymmetric magnetostatic equilibrium configurations. In particular,
it has been applied to obtain force-free magnetospheres around Kerr
black holes in the context of the energy extraction mechanisms for
relativistic jets by \citet{Blandford1977}. In their seminal work,
analytic solutions for the case in which the black hole (BH) spin is
small were obtained. More general solutions including arbitrarily
large values of the BH specific angular momentum require a numerical
evaluation of the solution of the GSE (e.g.,
\citealt{MacDonald:1984MNRAS.211..313,Fendt:1997A&A...319.1025,Uzdensky2004,Contopoulos2013,Nathanail2014}). As
an alternative to the solution of the GSE, the topology of the
electromagnetic field around a rotating BH has been determined as an
asymptotic steady state of force-free degenerate electrodynamics
(FFDE) evolution \citep{Komissarov2001,Komissarov2002, Komissarov2004,
 Tchekhovskoy2010}. \addnew{\citet{Tchekhovskoy2010} also construct
 steady state models for BH magnetospheres for a range of spin
 factors employing General Relativistic Magnetohydrodynamic (GRMHD)
 simulations in the force-free limit. These time-evolving approaches
 to reach a steady state usually impose boundary conditions at the
 outer BH event horizon as well as at the position of an assumed thin accretion disk.}

 Drawing
from previous findings on neutron star magnetospheres,
\cite{Contopoulos1999,Contopoulos2013} presented a numerical scheme for the solution
of the GSE in a split-monopole
setup. \citeauthor{Contopoulos2013} aimed to find a field line
configuration passing smoothly through the singular surfaces of the
problem (i.e., the light surfaces, as previously suggested by
\citealt{Lee2000}). For that, they
 implemented a numerical methodology relying vastly on subtle, empirically determined relaxation procedures of all
involved functions. In other words, the relaxation to the
 numerical solution requires recipes which seem to work, but there
 is no explicit mathematical justification about why they do. The
original algorithm of \cite{Contopoulos2013} has later been
improved in two ways \citep{Nathanail2014}. First, it has
 been supplemented by further smoothing steps in the
 numerical algorithm. Second, \cite{Nathanail2014} also included paraboloidal
configurations of the magnetic field. In their study of systems
of non-rotating BHs and thin accretion disks,
\cite{Uzdensky2004} found a solution to the GSE for a fixed field line angular velocity. He employs
the minimization of a suitably chosen error function at the light
surfaces in order to mathematically drive the numerical relaxation
procedure. A similar approach was followed by
\cite{Uzdensky:2005ApJ...620..889} in the case of rotating BHs
connected to thin accretion disks.

This paper begins by giving a recapitulation of the GSE and its
singular surfaces (sec.\,\ref{sec:GSmagentospheres}). With regards to
realistic field configurations in BH magnetospheres we formulate the
underlying equations of force-free electrodynamics in both the
potential and field representation. Subsequently
(sec.\,\ref{sec:numerics}), we present a comprehensive approach to the
numerical solution of the GSE. The strategy of minimizing a suitable
error function at the light surfaces (LS) \citep{Uzdensky2004} is
extended to the relaxation procedures of both, the field line angular
velocity as well as the current profile. We are able to quantify the
numerical errors and, hence, substantiate the quality and stability
of the found solutions. In sec.\,\ref{sec:NumericalResults}, the GSE
solution scheme is tested on split-monopole and paraboloidal
configurations\addnew{, as well as the test case of vertical magnetic fields}
\citep[cf.][]{Contopoulos2013,Nathanail2014}. Furthermore, a
current-free solution \citep[as found
in][]{Uzdensky:2005ApJ...620..889} is reproduced. The BZ process power
is studied for the split-monopole configurations in section
\ref{sec:BZpower}, emphasizing the need for reliable initial data of BH
magnetospheres with large spin parameter $a$.

\section{Grad-Shafranov equation for relativistic force-free Kerr
 magnetospheres}
\label{sec:GSmagentospheres}
The Kerr solution is a suitable approximation of the spacetime in
astrophysical scenarios of jet formation. It embodies the geometry of
a spinning BH of mass $M$ and specific angular momentum $a=J/M$ (with its dimensionless equivalent $a_*=a/M$), where
$J$ is the angular momentum. Throughout this work, the speed of light
and gravitational constant will be set as $c=G=1$. In Boyer-Lindquist
coordinates, the line element of the Kerr metric is
\begin{align*}
\text{d} s^2&=\:\left(1-\frac{2Mr}{\Sigma}\right)\text{d} t^2+\frac{4Mar\sin^2\theta}{\Sigma}\:\text{d} t\text{d} \phi\\
&-\frac{\Sigma}{\Delta}\:\text{d} r^2-\Sigma\:\text{d}\theta^2-\frac{A\sin^2\theta}{\Sigma}\:\text{d}\phi^2,
\end{align*}
\begin{align*}
\Sigma:=&\: r^2+a^2\cos^2\theta\:,\\
 A:=&\:\left(r^2+a^2\right)^2-\Delta\: a^2\sin^2\theta\:,\\
\Delta:= &\:r^2-2 M r+a^2 :=\left(r-r_+\right)\left(r-r_-\right)\:,\\
\end{align*}
where $r_\pm$ represent the locations
of the outer and inner horizons of the BH,
respectively. $r^*_\pm$ define the locations of the outer and inner ergosurfaces:
\begin{align}
r_\pm= M\pm\sqrt{M^2-a^2}\quad ; \quad r^*_\pm\addnew{(\theta)}=M\pm\sqrt{M^2-a^2\cos^2\theta}.
\end{align}
The frame-dragging frequency induced by the rotation of the BH is
 \begin{equation}
\Omega:=\:2aMr/A,
 \end{equation}
which is also the angular velocity of the (local)
 \textit{zero angular momentum observer} or ZAMO \citep[cf.][]{Thorne1986}, i.e.,
 $\Omega=(d\phi/dt)_{\rm ZAMO}$. 
 At the outer event horizon, the frame dragging frequency reads
\begin{align}
\OmegaBH:=\Omega(r=r_+) = \frac{a}{2Mr_+}=\frac{a}{r_+^2+a^2}.
\label{eq:BHRotation}
\end{align}
The redshift or lapse function is 
\begin{equation}
 \qquad \qquad \alpha:=\sqrt{\frac{\Sigma\Delta}{A}},
\end{equation}
 which accounts for the lapse of proper time $\tau$ in the
 ZAMO frame with respect to the global (Boyer-Lindquist) time $t$, thus,
 $\alpha=(d\tau/ dt)_{\rm ZAMO}$. While the global Boyer-Lindquist
 observer uses an spatial coordinate basis made by the set of
 orthogonal vectors $\{\partial_i\} = \{\boldsymbol{e}_i\}$, the
 local ZAMO observers have an attached tetrad
 $\{\boldsymbol{\hat{e}}_i\}=\{\boldsymbol{e}_i / \sqrt{g_{ii}}\}$,
 where the Latin index $i$ runs over the three spatial coordinates
 $(r,\theta,\phi)$. $g_{ii}$ are the diagonal components of the
 metric tensor, namely
\begin{align*}
g_{rr} = \frac{\Sigma}{\Delta}, \qquad g_{\theta\theta} = \Sigma, \qquad g_{\phi\phi}=\frac{A\sin^2{\theta}}{\Sigma}.
\end{align*}

The covariant
Maxwell equations governing the dynamics and topology of the
 electromagnetic field around a BH read
\begin{align}
 F^{\mu\nu}_{\:;\nu}=\:\addnew{\epsilon_0^{-1}}J^\mu \qquad\qquad ^{\ast}F^{\mu\nu}_{\:;\nu}=\: 0\:,
 	\label{eq:MaxwellCovariant}
 \end{align}
 where $F^{\mu\nu}$ and $^{\ast}F^{\mu\nu}$ are the Maxwell tensor and
 its dual, respectively, $J^\mu$ is the electric current four vector
 \addnew{and $\epsilon_0$ is the vacuum permittivity}. The semicolon
 denotes the covariant derivative. Since we seek time independent,
 force-balance configurations of the magnetosphere of a BH, we ignore
 the time derivatives involved in
 eq.\,(\ref{eq:MaxwellCovariant}). Under this assumption, the former
 set of equations can be cast in terms of 3-vectors measured by a ZAMO
 observer. Employing Boyer-Lindquist coordinates the former equations
 read \citep[cf.,][]{Thorne1986, Zhang:1989, Camenzind2007,
 Beskin2010}:

\begin{align}
 \boldsymbol{\nabla}\cdot\boldsymbol{\zamo{E}}=\;& 4\upi\rho\label{MaxIZamo}\:,\\
 \boldsymbol{\nabla}\cdot\boldsymbol{\zamo{B}}=\;& 0\:,\\
 \boldsymbol{\nabla}\times\left(\alpha\boldsymbol{\zamo{E}}\right)=&-\left(\boldsymbol{\zamo{B}}\cdot \boldsymbol{\nabla}\Omega\right)\cdot\boldsymbol{\hat{e}}_\phi\:,\\
 \boldsymbol{\nabla}\times\left(\alpha\boldsymbol{\zamo{B}}\right)=& -4\upi\alpha\boldsymbol{\zamo{j}}+\left(\boldsymbol{\zamo{E}}\cdot \boldsymbol{\nabla}\Omega\right)\boldsymbol{\hat{e}}_\phi\:,\label{MaxIVZamo}
\end{align}
where $\rho$, the 3-vectors $\boldsymbol{\zamo{E}}$,
$\boldsymbol{\zamo{B}}$ and $\boldsymbol{\zamo{j}}$ are the electric
charge density, the electric field, the magnetic field and the current
density measured by the ZAMO observer. $\boldsymbol{\hat{e}}_\phi$ is
the unit normal vector of the tetrad associated to the ZAMO in the
$\phi-$coordinate direction. In axisymmetric spacetimes it is
possible to distinguish between poloidal (along the potential lines
symmetric around the $\phi$-axis) and toroidal
($\boldsymbol{e}_\phi$-direction) components \citep[see
e.g.,][]{Punsly2001, Camenzind2007}.

To build up a stationary magnetosphere, it is necessary to guarantee
that there are either no forces acting on the system or, more
generally, that the forces of the system are in equilibrium. Except
along current sheets the latter condition
implies that the electric 4-current $J^{\mu}$ satisfies
the force-free condition \citep{Blandford1977}:
\begin{align}
F_{\mu\nu}J^\nu=0.
\label{eq:ForceFree}
\end{align}
Equation\,(\ref{eq:ForceFree}) is equivalent to a vanishing
Lorentz force on the charges in a the local ZAMO frame
\citep[see, e.g.,][]{Camenzind2007}:
\begin{align*}
\boldsymbol{E}\cdot\boldsymbol{j}=0\:,\\
\rho\boldsymbol{E}+\boldsymbol{j}\times\boldsymbol{B}=0\:.
\end{align*} 
These eqs. also imply the degeneracy condition
$\boldsymbol{E}\cdot\boldsymbol{B}=0$. Combining
eqs.\,(\ref{eq:MaxwellCovariant}) and (\ref{eq:ForceFree}) yields the
force-balance equation (or GSE) as introduced by
\cite{Blandford1977}. It relates the magnetic flux $\Psi(r,\theta)$
enclosed in the circular loop $r=\:$const., $\theta=\:$const. (divided
by $2\upi$) to the field line angular velocity
$\omega\left(\Psi\right)$ and the poloidal electric current
$I\left(\Psi\right)$ \citep[this version of the GSE is also used in,
e.g.,][]{Nathanail2014}:
\begin{align}
\begin{split}
4\frac{\Sigma}{\Delta}II'=&\left(\Psi_{,rr}+\frac{1}{\Delta}\Psi_{,\theta\theta}+\left(\frac{A_{,r}}{A}-\frac{\Sigma_{,r}}{\Sigma}\right)\Psi_{,r}-\frac{1}{\Delta}\frac{\cos\theta}{\sin\theta}\Psi_{,\theta}\right)\\
&\times\left[\frac{\omega^2 A \sin^2\theta}{\Sigma}-\frac{4Mar\omega\sin^2\theta}{\Sigma}-1+\frac{2 M r}{\Sigma}\right]\\
&+\left(\frac{A_{,r}}{A}-\frac{\Sigma_{,r}}{\Sigma}\right)\Psi_{,r}+\frac{4Mar\omega\sin^2\theta}{\Delta\Sigma}\frac{A_{,\theta}}{A}\Psi_{,\theta}\\
&-\frac{2Mr}{\Delta\Sigma}\frac{\Sigma_{,\theta}}{\Sigma}\Psi_{,\theta}\\
&+\left(2\:\frac{\cos\theta}{\sin\theta}+\frac{A_{,\theta}}{A}-\frac{\Sigma_{,\theta}}{\Sigma}\right)A\omega\left(\omega-\frac{4Mar}{A}\right)\frac{\sin^2\theta}{\Delta\Sigma}\Psi_{,\theta}\\
&-\left(\frac{2Mr}{\Sigma}-\frac{4Mar\omega\sin^2\theta}{\Sigma}\right)\left(\frac{A_{,r}}{A}-\frac{1}{r}\right)\Psi_{,r}\\
&+\frac{\sin^2\theta}{\Sigma\Delta}\left(A\omega-2Mar\right)\left(\Delta\:\omega_{,r}\:\Psi_{,r}+\omega_{,\theta}\:\Psi_{,\theta}\right),
\end{split}
\label{eq:GSLightCylinder}
\end{align}
The subscript comma indicates respective partial derivatives. From the
mathematical viewpoint, this equation is, in most of the space, an
elliptic, second-order partial differential equation (PDE) for the
magnetic flux \addnew{\citep[e.g.][]{Beskin:1997}}. This means that we
shall provide suitable boundary conditions to determine the solution
of the system. Since we are interested in employing the magnetospheric
configurations obtained with our new methodology as initial data for
evolutionary calculations, we shall compute the solution from the
outer event horizon of the BH to infinity. There is an added
complexity in the solution of the equation, since there are singular
surfaces of the spacetime, where the equation becomes a first order
PDE (see sec.\,\ref{sec:ligthcylinders}). Taking together these facts,
we shall devise a numerical method which adapts to the mathematical
(and physical) challenges in the type of PDE we have at hand.

A numerical solution to the GSE (eq. \ref{eq:GSLightCylinder}) will
consist of a relaxed configuration of the three functions
$\Psi(r,\theta)$, $\omega\left(\Psi\right)$ and
$I\left(\Psi\right)$. These functions fully determine the vector
fields $\left\{\boldsymbol{E},\boldsymbol{B}\right\}$ employed in
eq.\,(\ref{MaxIVZamo}) \citep[see, e.g.,][]{Camenzind2007}:
\begin{align}
	\boldsymbol{E}&=-\frac{\omega-\Omega}{2\upi\alpha}\boldsymbol{\nabla}\Psi\qquad\boldsymbol{B}_P=\frac{\boldsymbol{\nabla}\Psi\times\boldsymbol{e}_\phi}{2\upi\varpi^2}\qquad B^T=-\frac{2I}{\alpha\varpi^2}.
	\label{eq:PotFieldConversion}
\end{align}
Here, $\varpi=\sqrt{-g_{\phi\phi}}$ is the cylindrical radius,
$\boldsymbol{B}_P$ represents the poloidal magnetic field and $B^T$
the toroidal magnetic field component. In their field representation,
solutions to the GSE will eventually be employed in conservative time
evolution schemes of force-free electrodynamics \citep[as suggested,
e.g., by][]{Komissarov2004,Komissarov2007}.

\subsection{Light surfaces}
\label{sec:ligthcylinders}

The numerical solution of the GSE relies on the use of additional
regularity conditions at the singular surfaces of
eq.\,(\ref{eq:GSLightCylinder}). Throughout the domain, the so called
light surfaces (LS) are situated where the coefficient multiplying the
second order derivatives vanishes, i.e., where the condition
\begin{align}
\addnew{\mathcal{D}:=\frac{\omega^2 A \sin^2\theta}{\Sigma}-\frac{4Mar\omega\sin^2\theta}{\Sigma}-1+\frac{2 M r}{\Sigma}=0}
\label{eq:LCCondition}
\end{align}
is satisfied. In an analogy to the pulsar magnetosphere
\citep{Ruderman1975}, the LS can be understood as singular surfaces
where magnetic field lines rotate superluminally with respect to the
ZAMO observer (e.g., \citealt{Komissarov2004}). In that context they
are known as light cylinders. Outside of the outer light surface
(OLS), magnetic field lines rotate faster than the speed of light with
respect to ZAMOs. Inside the inner light surface (ILS), magnetic field
lines counterrotate superluminally with respect to the ZAMO. The ILS
falls inside the ergosphere and touches its boundary (and, hence, also
the outer horizon) at the rotational axis of the system (located at
$\theta=0$). As explicitly shown in \cite{Komissarov2004}, while the
radial coordinate $r$ of the ILS increases monotonically with $\theta$
between the rotational axis and the equator, the opposite holds for
the OLS.

Across these singular surfaces we demand regularity of the three
scalar functions $\Psi\left(r,\theta\right)$,
$\omega\left(\Psi\right)$ and $I\left(\Psi\right)$. More specifically,
we require that the magnetic flux function $\Psi$ crosses smoothly
through the ILS and through the OLS. The remaining two functions
$\omega\left(\Psi\right)$ and $I\left(\Psi\right)$ will be
reconstructed from the smooth $\Psi$ function. If condition
(\ref{eq:LCCondition}) holds, then eq.~(\ref{eq:GSLightCylinder})
becomes the \emph{reduced} GSE, which allows to relate the
aforementioned three functions through
\begin{align}
\begin{split}
4\frac{\Sigma }{\Delta}II'=&\left(\frac{A_{,r}}{A}-\frac{\Sigma_{,r}}{\Sigma}\right)\Psi_{,r}-\frac{2Mr}{\Delta\Sigma}\frac{\Sigma_{,\theta}}{\Sigma}\Psi_{,\theta}+\frac{4Mar\omega\sin^2\theta}{\Delta\Sigma}\frac{A_{,\theta}}{A}\Psi_{,\theta}\\
&+\left(2\:\frac{\cos\theta}{\sin\theta}+\frac{A_{,\theta}}{A}-\frac{\Sigma_{,\theta}}{\Sigma}\right)A\omega\left(\omega-\frac{4Mar}{A}\right)\frac{\sin^2\theta}{\Delta\Sigma}\Psi_{,\theta}\\
&-\left(\frac{2Mr}{\Sigma}-\frac{4Mar\omega\sin^2\theta}{\Sigma}\right)\left(\frac{A_{,r}}{A}-\frac{1}{r}\right)\Psi_{,r}\\
&+\frac{\sin^2\theta}{\Sigma\Delta}\left(A\omega-2Mar\right)\left(\Delta\:\omega_{,r}\:\Psi_{,r}+\omega_{,\theta}\:\Psi_{,\theta}\right).
\end{split}
\label{eq:GSReduced}
\end{align}
As noted by \cite{Uzdensky:2005ApJ...620..889}, the reduced GSE must
be fulfilled, both, at the ILS and at the OLS. Thus, we have two
relations among the freely specifiable functions $\omega(\Psi)$ and
$I(\Psi)$.

\section{A generalized numerical Grad-Shafranov solver}
\label{sec:numerics}
\begin{figure}
	\centering
	\includegraphics[width=0.49\textwidth]{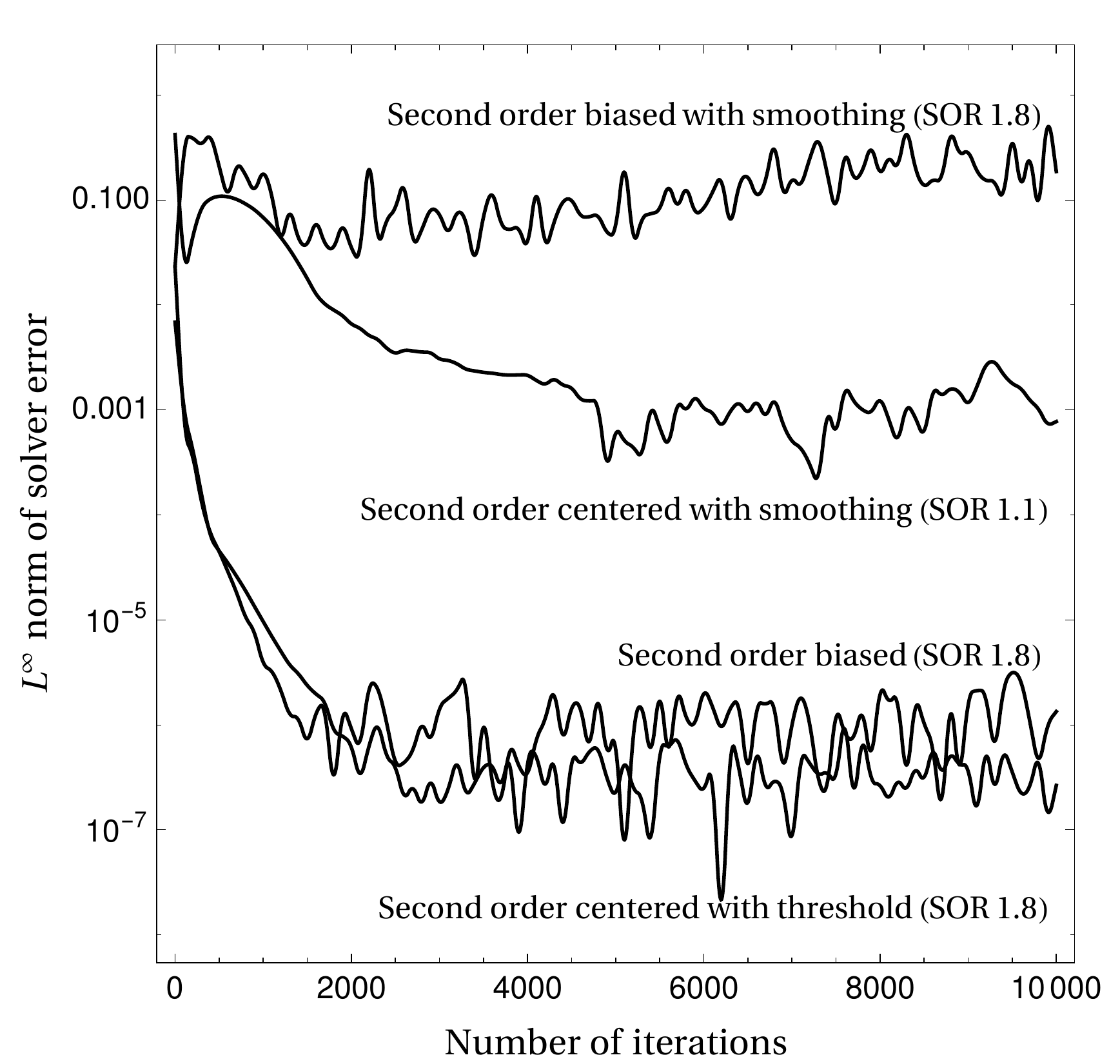} 
	\caption{Evolution of the $L^\infty$ norm of the solver
 residual (cf. eq. \ref{eq:SolverRes}). Comparison of
 different discretization schemes and their convergence
 behavior inside the OLS ($[r_+,3.0] \times [0,90^\circ]$,
 $\left[n_r\times n_\theta\right]=\left[200\times
 100\right]$) for a BH with $a_*=0.9999$, during
 relaxation of $II'\left(\Psi\right)$ for
 $\omega\left(\Psi\right)=0.5\Omega_{BH}$ fixed to the
 initial value. The initial magnetic flux distribution
 corresponds to that of a split-monopole (see
 sec.\,\ref{sec:smconfigs}). Relaxation coefficients of the
 successive overrelaxation (SOR) scheme are chosen according
 to maximal convergence without numerical breakdown of the
 iterative scheme. Their values are written in parenthesis
 for each different case. The presented tests consist of (i)
 a second order finite difference discretization with smoothing at he LS
 in every iteration, (ii) a second order finite difference
 discretization with a threshold on the coefficients
 $\mathcal{C}_{rr}$ and $\mathcal{C}_{\theta\theta}$ ensuring
 diagonal dominance of eq. (\ref{eq:CoeffMat}), (iii) the second
 order discretization with biased stencil at the LS and
 additional smoothing in every step, and (iv) the second
 order discretization with biased stencil at the LS with no
 additional smoothing.}
	\label{fig:ConvergenceL}
\end{figure}
\begin{figure}
	\centering
	\includegraphics[width=0.49\textwidth]{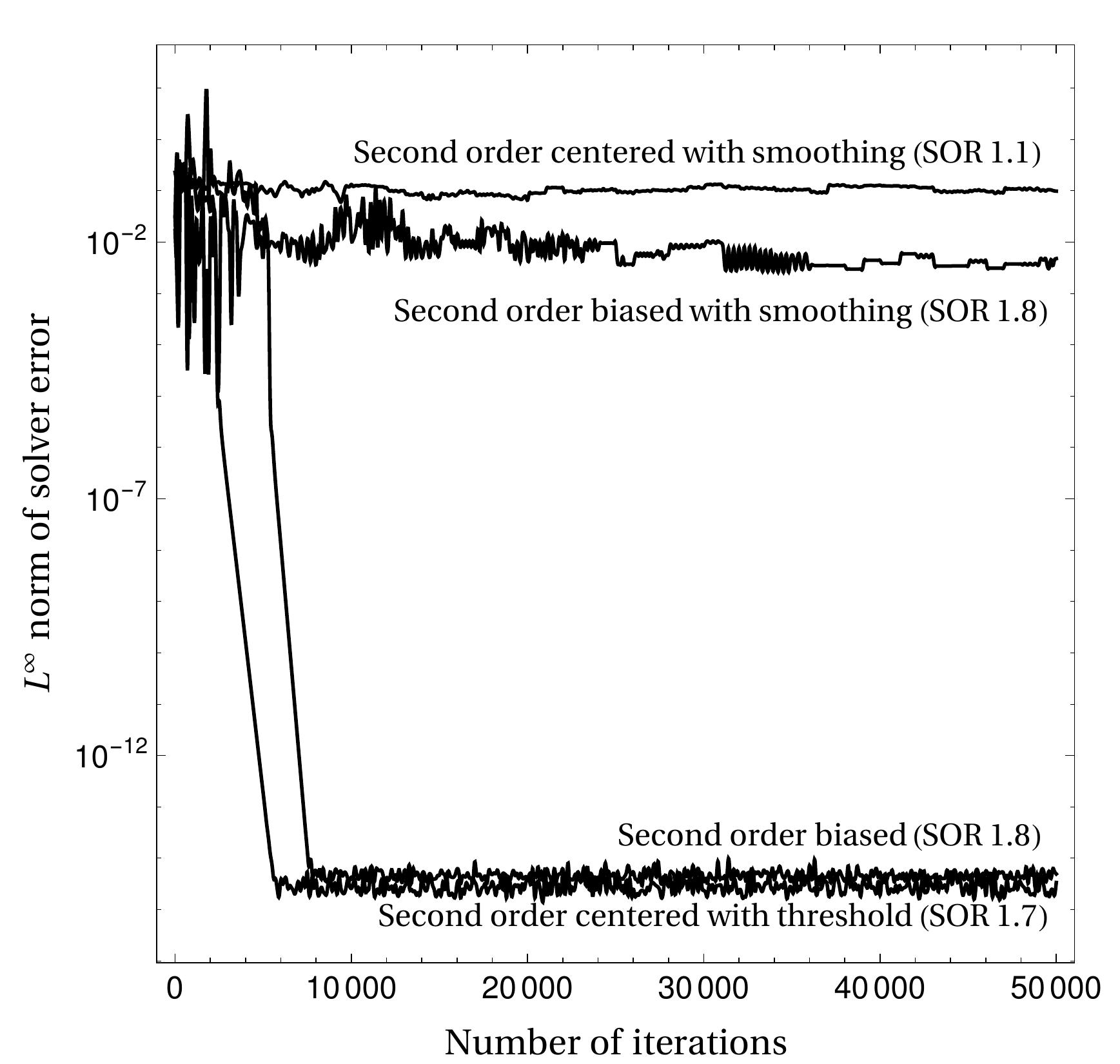} 
	\caption{Evolution of the $L^\infty$ norm of the solver
 error. Comparison of different discretization schemes and
 their convergence behavior including the ILS and the OLS
 ($[r_+,\infty] \times [0,90^\circ]$,
 $\left[n_r\times n_\theta\right]=\left[200\times
 100\right]$) for a BH with $a_*=0.9999$, during
 relaxation of both $II'\left(\Psi\right)$ and
 $\omega\left(\Psi\right)$. The iterative procedure on
 $\omega$ and $II'$ proceeds as long as
 $\mathcal{R}_{\Psi}>10^{-5}$. SOR factors are chosen
 according to maximal convergence without numerical breakdown
 of the iterative scheme. Their values are written in
 parenthesis for each different case. The initially guessed
 magnetic flux distribution corresponds to that of a
 split-monopole (see sec.\,\ref{sec:smconfigs}).}
	\label{fig:ConvergenceLLong}
\end{figure}
\begin{figure}
	\centering
	\includegraphics[width=0.49\textwidth]{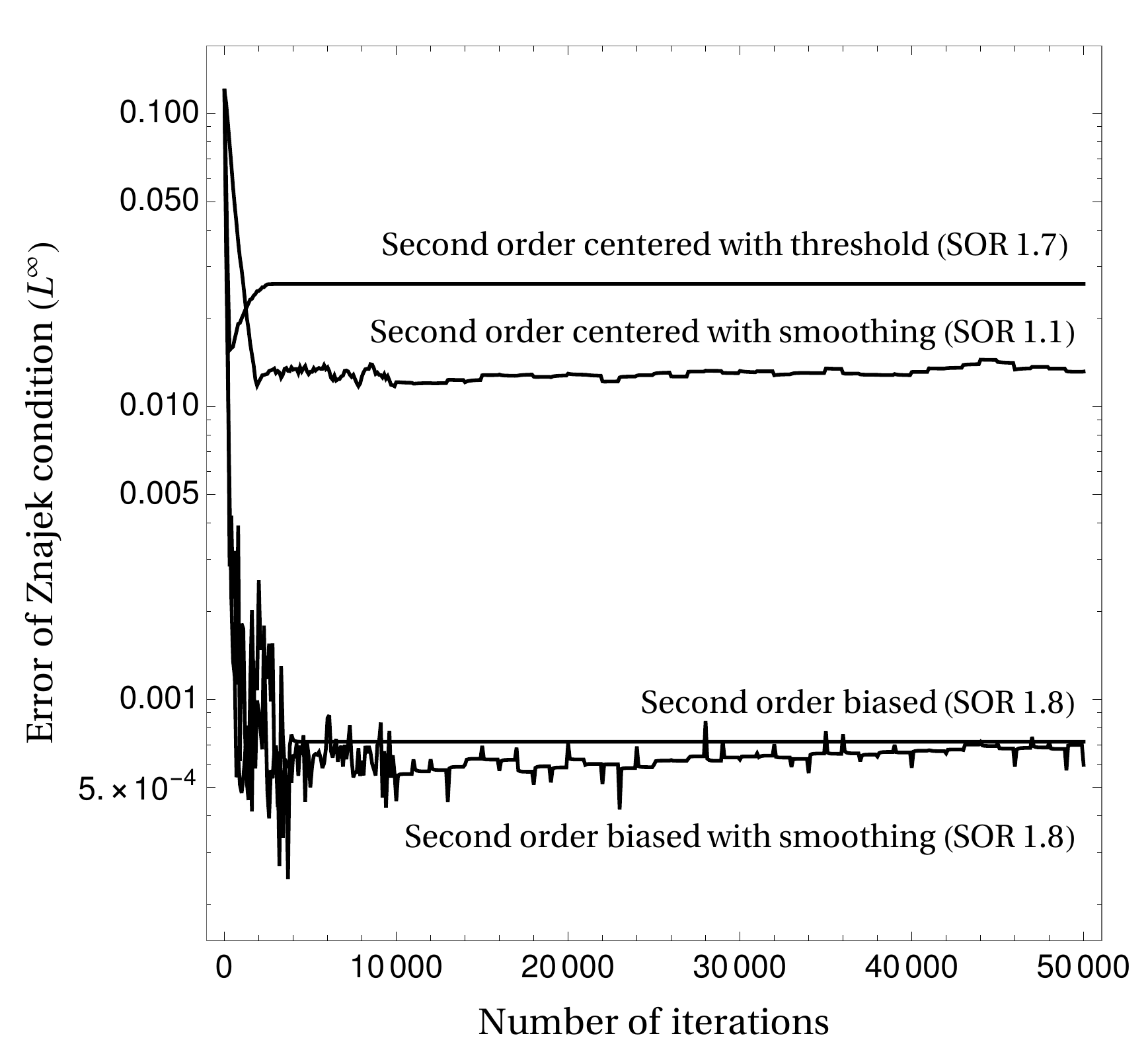} 
	\caption{Same problem setup and numerical
 methodologies as in fig.\,(\ref{fig:ConvergenceLLong}) but
 here showing the evolution of the $L^\infty$ norm of the
 deviation from the Znajek condition at the horizon. The
 second order biased stencil provides both fast convergence
 and evolution towards a configuration fulfilling the
 Znajek condition. }
	\label{fig:ConvergenceZ}
\end{figure}


Our method is based on a finite-difference solution of
eq. (\ref{eq:GSLightCylinder}). For that, we discretize all the
physical and geometrical quantities in a two dimensional grid. The
radial coordinate is compactified according to the transformation
$R\left(r\right)=r/\left(r+M\right)$ as introduced by
\citet{Contopoulos2013}. Radial derivatives are mapped to the
$R\left(r\right)$ coordinate by the following transformations:
\begin{align}
\begin{split}
\frac{\partial}{\partial r}=&\:\left[\frac{M}{\left(r+M\right)^2}\right]\frac{\partial}{\partial R}\\
\frac{\partial^2}{\partial r^2}=&\: -\left[\frac{2}{\left(r+M\right)^3}\right]\frac{\partial}{\partial R}+\left[\frac{M}{\left(r+M\right)^2}\right]^2\frac{\partial^2}{\partial R^2}
\end{split}
\label{eq:CoordChangeR}
\end{align}
The computational domain covers the region
$[R_{\rm min}, R_{\rm max}] \times [0,\theta_{\rm max}]$, where
$R_{\rm min}=r_{+}/\left(r_{+}+M\right)$ (i.e., the computational
domain extends radially down to the outer event horizon) and
$R_{\rm max}$ is specified differently according to the application we
seek. The region mapped by the grid may easily be extended to reach
all the way to infinity at $R_{\rm max}=1$. In most cases we set
$\theta_{\rm max}=\upi/2$ and symmetry with respect to the equatorial
plane. Given that the Kerr metric fulfills this property, it is
reasonable to search for solutions of the GSE with this symmetry as
well. The $\left(R,\theta\right)$ domain is covered by a uniform
mesh, where the number of mesh points in the $r$ and $\theta$
directions is $n_r$ and $n_\theta$, respectively. The discrete values
of the magnetic flux $\Psi_{ij}:=\Psi(R(r_i),\theta_j)$ are stored on
a two-dimensional array of the same size as the numerical grid,
whereas the two remaining functions $\omega\left(\Psi\right)$ and
$II'\left(\Psi\right)$ are tabulated as a one-to-one map of $\Psi$.
For practical purposes, instead of working directly with the function
$I(\Psi)$, we use $II'(\Psi)$ \citep[cf.][]{Nathanail2014}. The latter
is related to the former by
\begin{align*}
|I\left(\Psi\right)|=\left[2\int_{0}^{\Psi}II'\left(\Psi\right)\right]^{1/2}\;.
\end{align*}
One should note that the additional arbitrariness of sign induced by the
prescribed recovery of the current $I\left(\Psi\right)$ from the
function $II'\left(\Psi\right)$ should be handled carefully in
eq.\,(\ref{eq:PotFieldConversion}). The numerical solution
determining $\Psi$, $\omega\left(\Psi\right)$ and $II'(\Psi)$ is
obtained using an iterative procedure. In this iteration, the initial
values of these functions can be specified freely.

Physically the magnetosphere is divided into three disconnected
regions by the two light surfaces of the problem. Mathematically, we
shall map this property by solving \emph{independently} for the scalar
function $\Psi$ in each subdomain. The only connection between domains
are the regularity conditions at the separatrices among
subdomains. Accounting for these facts, the numerical method we
propose splits each iteration into three basic blocks of (1) the
finite difference solution of the GSE in each of the subdomains, (2)
the matching of the solutions across the light surfaces to obtain
regular functions and (3) the build-up or update of the functional
tables for $\omega(\Psi)$ and $II'(\Psi)$. In the following sections,
the details of each of these blocks are provided.

\subsection{Finite difference solution of the GSE in each subdomain}
\label{sec:finiteDifSol}

For the finite difference solution of the GSE in each subdomain we
take advantage of the existing computational infrastructure for
linear elliptic PDEs used in \cite{Adsuara2015}. In order
to apply these methods to the non-linear equation at hand, we split
the GSE into a term linear in the derivatives of $\Psi$ (right-hand
side of eq.\,\ref{eq:GSLightCylinder}) and into a part comprising the
non-linear source terms (left-hand side of
eq.\,\ref{eq:GSLightCylinder}). The coefficients of the derivatives as
well as the source terms are discretized on the mesh. The GSE
(\ref{eq:GSLightCylinder}) can be written \addnew{in canonical form} as
\begin{align}
\mathcal{C}_{rr}\Psi_{,rr}+\mathcal{C}_{\theta\theta}\Psi_{,\theta\theta}+\mathcal{C}_{r}\Psi_{,r}+\mathcal{C}_{\theta}\Psi_{,\theta}=\mathcal{S},
\label{eq:canonicalform}
\end{align}
where
$\left(\mathcal{C}_{r},\mathcal{C}_{rr},\mathcal{C}_{\theta},\mathcal{C}_{\theta\theta}\right)$
are the PDE coefficients and $\mathcal{S}$ the sources (left-hand side
of eq.\,\ref{eq:GSLightCylinder}). \addnew{We note that the GSE is
 linear in the higher order derivatives, and that it contains no
 terms proportional to $\Psi_{,r\theta}$,
 i.e. $\mathcal{C}_{r\theta}=0$. Following, e.g. \cite{Beskin:1997},
 it is then easy to see from the canonical form of the GSE
 (eq.\,\ref{eq:canonicalform}), that the character of the equation
 depends on the sign of the discriminant
 $\mathcal{C}_{r\theta}^2-4
 \mathcal{C}_{rr}\mathcal{C}_{\theta\theta}=-4\mathcal{D}^2/\Delta$. Since
 $\Delta>0$ for $r>r_+$ and $\mathcal{D}^2>0$ everywhere except at
 the LSs, the GSE is elliptic. At the LSs ($\mathcal{D}=0$) the
 character of the equation does not change because of the regularity
 condition given in eq.\,\ref{eq:GSReduced}.}

Employing a second order centered
finite difference scheme on an equally spaced grid, the discretized
form of the GSE reads
\begin{align}
\begin{split}
\mathcal{S}_{i,j}=\:
&\Psi_{i-1,j}\left[\frac{\mathcal{C}_{rr}}{\delta r^2}-\frac{\mathcal{C}_{r}}{2\delta r}\right]-\Psi_{i,j}\left[\frac{\mathcal{C}_{rr}}{\delta r^2}+\frac{\mathcal{C}_{\theta\theta}}{\delta \theta^2}\right]\\
&+\Psi_{i+1,j}\left[\frac{\mathcal{C}_{rr}}{\delta r^2}+\frac{\mathcal{C}_{r}}{2\delta r}\right]+\:\Psi_{i,j-1}\left[\frac{\mathcal{C}_{\theta\theta}}{\delta\theta^2}-\frac{\mathcal{C}_{\theta}}{2\delta \theta}\right]\\
&+\Psi_{i,j+1}\left[\frac{\mathcal{C}_{\theta\theta}}{\delta \theta^2}+\frac{\mathcal{C}_{\theta}}{2\delta \theta}\right],
\end{split}
\label{eq:DiscreteGSE}
\end{align}
where we have dropped subscripts $(i,j)$ of the coefficients
$\left(\mathcal{C}_{r},\mathcal{C}_{rr},\mathcal{C}_{\theta},\mathcal{C}_{\theta\theta}\right)$
to avoid cluttering the formulae with subindices. From this
discretization, a coefficient matrix is built and used for the
iterative relaxation procedure:

 \begin{align}
{\cal G}_{S}:=
{\footnotesize \left( 
\arraycolsep=1.4pt\def\arraystretch{1.5}
\begin{array}{ccccccc}
\II{c}	& \II{u} & \zze &\cdots& \cdots&\cdots& \zze \\ 
\II{d} & \II{c} & \II{u} & \zze & \cdots&\cdots & \zze \\ 
\zze & \II{d} & \II{c} & \II{u} & \zze & \cdots& \zze \\ 
\vdots & \zze &\ddots&\ddots&\ddots& \vdots & \zze \\ 
\vdots & \vdots &\ddots&\ddots&\ddots& \ddots & \zze \\ 
\vdots & \vdots& \vdots&\vdots &\II{d} & \II{c} & \II{u} \\ 
\zze & \cdots & \cdots& \cdots & \zze &\II{d} & \II{c}\\ 
\end{array}\right)},
\label{eq:CoeffMat}
\end{align}
Here, $\II{c}$, $\II{u}$ and $\II{d}$ are matrices with dimensions
$n_r \times n_\theta$, which contain the combinations of coefficients
of eq.\,(\ref{eq:DiscreteGSE}) and $\zze$ is the null matrix with
dimensions $n_r \times n_\theta$.

The numerical elliptic PDE solver is used with an iterative SOR
(successive overrelaxation) scheme to find the magnetic flux function
$\Psi$. For the complex non-linear system at hand, there is no known
optimal relaxation coefficient of the SOR scheme,
$\omega_{\rm SOR, opt}$. Thus, we need to choose a value
$\omega_{\rm SOR}$ empirically. Numerical experience tells that we
shall take a value as close as possible to 2, but not too large such that
the iterative scheme diverges. The choice of $\omega_{\rm SOR}$
strongly depends upon the grid properties (e.g., number of grid
points, physical domain size) as well as the numerical treatment of
the LS.

Both, the grid extension and the discretization stencil have an impact
on the diagonal dominance of the resulting coefficient matrices
(eq.\,\ref{eq:CoeffMat}) of the
solver. 
In case of the relativistic GSE (eq.\,\ref{eq:GSLightCylinder}),
diagonal dominance may be greatly breached at the location of the
singular surfaces (cf. condition \ref{eq:LCCondition}), where the
coefficients $\mathcal{C}_{rr}$ and $\mathcal{C}_{\theta\theta}$
vanish. This is mostly due to the fact that points across a separatrix
of the computational domain should not be bridged by the finite
difference discretization. Stated differently, a derivative on a given
computational subdomain must not include values on a different
subdomain in its stencil. We point out that this fact was brought
about by \cite{Camenzind:1987A&A...184..341}, but in the context of
the finite element solution of the GSE.
\cite{Camenzind:1987A&A...184..341} points out, that, as the finite
element grid must follow the shape of the light surfaces, the nodal
points had to be redistributed iteratively in his numerical
method. Turning to our finite difference discretization, we shall see
that, e.g., a standard second order centered finite difference
approximation of the first derivatives $\Psi'$ couples points across
LS, rendering a poor convergence (if at all) to the solution. This
fact forces us to employ $\omega_{\rm SOR}$ closer to 1, instead to
2. Changing the discretization for the cells around the LS to a
left/right biased second order scheme or reducing the approximation to
first order of accuracy greatly improves diagonal dominance of the
coefficient matrix and, hence, convergence behavior of the numerical
solver (see fig.\,\ref{fig:ConvergenceL}).

If not stated otherwise, Dirichlet boundary conditions are imposed
along the symmetry axis as well as on the equator in the simulations,
where we fix the minimum and maximum values of the potential $\Psi$,
respectively. Newman boundary conditions are set up along the radial
edges of the computational domain. The latter implies that we set up
the derivatives of $\Psi$ normal to the outer horizon at $r=r_+$. Note
that the value of $\Psi$, or of any other free function of $\Psi$, is
not imposed at the outer event horizon. In particular, the so called
Znajek condition \citep{Blandford1977,Znajek1977} is not explicitly
enforced there.

The iterative solution is stopped when we attain a prescribed
reduction of the residual, defined as
 \begin{align}
\mathcal{R}_\Psi=\left|\Psi^{(n)}-\Psi^{(n-1)}\right|_\infty ,
\label{eq:SolverRes}
\end{align}
where $|.|_\infty$ stands for the $L^\infty$ norm computed over all
the discrete points of our numerical grid (for more details see
app. \ref{sec:convergence}).

\subsection{Matching across subdomains}
\label{sec:Matching}
%
To ensure regularity of the potential $\Psi$ across the light
surfaces, we have employed two strategies. First, we perform a cycle
consisting of iterative overrelaxations of the GSE interleaved with
numerical resets of the values of $\Psi$ developed at the light
surfaces. The mentioned cycle starts computing a series of iterations
of the solution on each of the three subdomains independently. This
brings a mismatch between solutions across subdomains. The most
severe mismatch happens at the ILS, where numerical artifacts
develop. In order to smooth out the solution, we build high-order
Lagrange interpolation polynomials in the radial direction for
$\Psi$. These polynomials have a stencil centered around the light
surfaces on each different discrete value of $\theta_j$
($j=1,\ldots,n_\theta$). Thus, they encompass points in two different
computational domains. At the radial location of the light surfaces we
obtain an smooth interpolant of $\Psi$, which replaces the numerical
values (artifacts) developed there in the course of the iterative
solution. We repeat the whole cycle until convergence is reached. This
first strategy follows from \cite{Nathanail2014}, but we employ
higher-order polynomial interpolants for $\Psi$ ($5^{th}$ order
Lagrangian interpolation, instead of just taking for $\Psi$ the
average between its values on both sides of a light surface -
cf. eq.\,(15) of \citealt{Nathanail2014}).

The second strategy consists in producing a central, second order
finite difference discretization in all points of the computational
domain except close to the light surfaces. There we switch to a
(left/right) biased, second order, finite difference discretization of
the first derivatives of the GSE. This procedure notably reduces the
coupling between different physical domains. However, since the light
surfaces are not spherical, some unwanted couplings may develop due to
the discretization of angular derivatives. As a result of the biased
discretization the coefficient matrix of the linear system to be
solved (eq.\,\ref{eq:CoeffMat}) improves its diagonal dominance. The
improved diagonal dominance results in a faster convergence of the
method than when no biased discretizations are employed (as we shall
see in sec.\,\ref{sec:NumericalResults}). In
fig.\,\ref{fig:ConvergenceL} we clearly see that a second order biased
discretization around the light surfaces works better if no smoothing
is applied to $\Psi$. Indeed, with the use of a biased discretization
the need of any smoothing of the solution at the light surfaces disappears and we do not apply
it. \addnew{This matching strategy follows the general
 guidelines devised by \cite{LeVeque:1994} for the treatment of
 immersed boundaries in second-order elliptic equations.}

A third strategy has also been tested, namely, we employ a second
order centered discretization everywhere, but at the light surfaces we
use a threshold for the coefficients $\mathcal{C}_{rr}$ and
$\mathcal{C}_{\theta\theta}$ ensuring diagonal dominance of the matrix
of the system (eq.\,\ref{eq:CoeffMat}). Note that
$\mathcal{C}_{rr}=\mathcal{C}_{\theta\theta}=0$ on the light
surfaces. Thus, the proposed recipe consists of replacing the
aforementioned coefficients by
\begin{align*}
\mathcal{C}_{rr} &= \textrm{sign}(\mathcal{C}_{rr}) \times \max{(|\mathcal{C}_{rr}|, \epsilon)}, \\
\mathcal{C}_{\theta\theta} &= \textrm{sign}(\mathcal{C}_{\theta\theta}) \times \max{(|\mathcal{C}_{\theta\theta}|, \epsilon)} ,
\end{align*}
with $\epsilon\sim 10^{-5}$. As in the case of the second strategy,
the thresholding of the coefficients of the second order derivatives
renders unnecessary any smoothing procedure at the light
surfaces. Fig.~\ref{fig:ConvergenceL} shows that the second and third
strategies yield a quite similar reduction of the residual with the
number of iterations.

In fig.\,\ref{fig:ConvergenceLLong} we show the evolution of the
residual with the number of iterations in the solver, again, for
different matching strategies. Differently from
fig.\,\ref{fig:ConvergenceL}, in this case we include the whole space
time ($r_{\rm max}=\infty$). Regardless of whether we set the outer
boundary conditions at finite or infinite distance, the qualitative
conclusion is the same. Namely, either thresholding the coefficients
of the second order derivatives, or employing a biased discretization
close to the light surfaces brings a much larger reduction (by roughly
9 orders of magnitude) than smoothing the solution across the light
surfaces. Furthermore, smoothing procedures are unable to reduce
substantially the residual for coarse discretizations. The results
shown in figs. \ref{fig:ConvergenceLLong} and \ref{fig:ConvergenceL}
also hold for higher resolutions.

Since the second strategy presented in this section (left/right biased
stencils) does not depend on any additional tunable parameter and
since it yields a reduction of the residual comparable to the case of
using thresholding, we will use it as our default method to match the
solution across different subdomains.

\subsection{Update of the potential functions}
\label{sec:UpdatesPotentials}

The potential functions $\omega\left(\Psi\right)$ and/or
$II'\left(\Psi\right)$ could be updated every time the magnetic
flux $\Psi$ changes in the course of the iterative relaxation sketched
in sec.\,\ref{sec:finiteDifSol}. In practice, it is unnecessary to
update $\omega\left(\Psi\right)$ and $II'\left(\Psi\right)$ with this
frequency. Instead, the mentioned update is performed after
$n_{\rm u}\ge 1$ iterations. The choice of $n_{\rm u}$ comes as a
tradeoff between accuracy and computational time.

The update of both functions simultaneously
\citep[see][]{Contopoulos2013}, as well as with one of them fixed
\citep[cf.][]{Uzdensky2004} to an initially specified value are
equally possible in our scheme. For convergence testing we have
considered both cases, i.e., the relaxation of either
$\omega\left(\Psi\right)$ (not shown here) or
$II'\left(\Psi\right)$ (fig.\,\ref{fig:ConvergenceL}) and of both functions
simultaneously (fig.\,\ref{fig:ConvergenceLLong}). A cautionary note
must be added here. The number of light surfaces in the computational
domain determines whether one or none of the potential functions can
be arbitrarily set up. More precisely, the number of freely
specifiable potential functions equals two minus the number of light
surfaces in the domain. For instance, if the OLS radius is
sufficiently large (e.g., when $a\rightarrow 0$), the outermost radial
computational domain may be set inside of the OLS for numerical
convenience. In this case, we are allowed to freely specify either
$\omega(\Psi)$ or $I(\Psi)$. This is
the simplification we employ to obtain the results shown in
fig.\,\ref{fig:ConvergenceL}. Note, however, that if the numerical
domain contains both LS, there is no freedom to set the potential
functions. They must be recovered from eq.\,(\ref{eq:RLC}) applied
both at the ILS and the OLS. The convergence properties of the latter
case can be seen in fig.\,\ref{fig:ConvergenceLLong}. In view of the
results, the global convergence properties of the algorithm are not
sensitively dependent on the choice of updating only one or both
potential functions.

The updates of the potential functions are conducted by minimizing the
error of eq.\,(\ref{eq:GSReduced}) after determining the exact radial
position of the LS and the corresponding interpolated
quantities. More specifically, we define the residual at the LS as
\begin{align}
\begin{split}
\mathcal{R}_{\rm LC} = &\left|\: 4\frac{\Sigma }{\Delta}II' - \right. 
\left(\frac{A_{,r}}{A}+\frac{\Sigma_{,r}}{\Sigma}\right)\Psi_{,r}-\frac{2Mr}{\Delta\Sigma}\frac{\Sigma_{,\theta}}{\Sigma}\Psi_{,\theta}\\ &-\frac{4Mar\omega\sin^2\theta}{\Delta\Sigma}\frac{A_{,\theta}}{A}\Psi_{,\theta}\\
&-\left(2\:\frac{\cos\theta}{\sin\theta}+\frac{A_{,\theta}}{A}-\frac{\Sigma_{,\theta}}{\Sigma}\right)A\omega\left(\omega-\frac{4Mar}{A}\right)\frac{\sin^2\theta}{\Delta\Sigma}\Psi_{,\theta}\\
&+\left(\frac{2Mr}{\Sigma}-\frac{4Mar\omega\sin^2\theta}{\Sigma}\right)\left(\frac{A_{,r}}{A}-\frac{1}{r}\right)\Psi_{,r}\\
&\left.-\frac{\sin^2\theta}{\Sigma\Delta}\left(A\omega-2Mar\right)\left(\Delta\:\omega_{,r}\:\Psi_{,r}+\omega_{,\theta}\:\Psi_{,\theta}\right)
\: \right|,
\end{split}
\label{eq:RLC}
\end{align}
and attempt to minimize it (see also the convergence criterion in
app.\,\ref{sec:convergence}). The process of minimizing $R_{\rm LC}$
depends on whether we fix one of the two free potential functions (and
which one of them) or if we leave both to be numerically obtained from
the reduced GSE (eq.\,\ref{eq:GSReduced}) applied at both
LS. Independent of the fixing or relaxing of the function
$\omega(\Psi)$ (see below), the functional update of $II'$ is achieved
in a straightforward manner by substitution of $\omega$ and $\omega'$
into the right-hand side of expression (eq.\,\ref{eq:GSReduced}) at
the LS. If we do not initially specify the rotational profile and keep
it throughout the iterative solution, then we need to provide initial
guesses $\omega_{0}$ and $\omega'_{0}$ for $\omega$ and its
derivative, respectively. We note that every time $\omega$ is changed,
the location of the LS (eq.\,\ref{eq:LCCondition}) changes. The
practical procedure consists on taking a set of a few thousands of
values $\omega_{0}$ and $\omega'_{0}$ uniformly selected in the
intervals $[\omega(\Psi) - \xi, \omega(\Psi) + \xi]$ (throughout the
shown tests we use $\xi=0.15$) and for each of these values we
compute $R_{\rm LC}$ (eq.\,\ref{eq:RLC}). Among all these pairs of
values $\omega_{0}$ and $\omega'_{0}$ we pick the one which minimizes
$R_{\rm LC}$.

Optimal and stable results require an exact localization of the LS
positions. Since we employ a finite difference method, the spatial
discretization determines the numerical accuracy with which the
singular surfaces are resolved. However, for practical grid
resolutions, it is necessary to exceed the accuracy of the numerical
grid in order to achieve high accuracy in resolving
eq. (\ref{eq:RLC}). Once it is detected that a given numerical cell
between grid points, namely, bounding the region
$[r_i,r_{i+1}]\times[\theta_j,\theta_{j+1}]$, is traversed by a LS,
our algorithm improves the accuracy of its localization using either
Lagrangian interpolation polynomials or bicubic spline interpolation
if a higher resolution is desired. In the presented solutions,
especially for lower values of the BH spin parameter $a$, the
numerical grid is refined before every update of the potential
functions and bicubic spline interpolation is used to determine the
quantities at the respective LS. For small values of $\theta$ (i.e.,
the first few zones at the $\theta_{min}$ boundary), we approximate
the potential $\Psi$ by the initially guessed function in order to
avoid numerical artifacts due to the small distance between the ILS
and the event horizon \citep[as suggested by][]{Contopoulos2013}.

The functions $\omega_{\rm ns}(\Psi)$ and $II'_{\rm ns}(\Psi)$
obtained with the previous procedures tend to be non-smooth. This lack
of smoothness degrades the convergence properties of the finite
difference solution. Thus, we replace $\omega_{\rm ns}$ and
$II'_{\rm ns}$ by smooth cubic spline interpolants of the latter
functions \citep[cf.][]{Contopoulos2013,Nathanail2014}. For that, we pick a sample of $n_{\rm int}$ values of both
$\omega_{\rm ns}$ and $II'_{\rm ns}$ as nodal points for the cubic
spline interpolation. The number of nodal points may be chosen in the
algorithm setup and may influence the accuracy of the solution (the
presented runs employed $n_{\rm int, \omega}=5$ and
$n_{\rm int, II'}=10$). Especially for the first relaxation steps a
lower order of $n_{\rm int, \omega}$ and $n_{\rm int, II'}$ may be
beneficial in order to prevent undesired oscillations. The presented
procedure has been tested as well with $n_{\rm int, \omega}=10$ and
$n_{\rm int, II'}=20$ differing in the rate of initial convergence
without noticeable changes to the relaxed solution.

\cite{Uzdensky2004} has applied the update of the current function
employing eq.\,(\ref{eq:GSReduced}) in order to find the field
configuration of a central engine with the field line angular velocity
fixed by the disk's rotation. With the suggested methodology for the
updating of the potential functions, we generalize his approach by
allowing the fixing of either potential function.

\subsection{The Znajek condition at $r_+$}
%
A key ingredient of the derivation of an outflowing energy and a
process efficiency measure at the horizon in \cite{Blandford1977} is
the so called Znajek boundary condition. Historically by
\cite{Weber1967} and context specific by \cite{Znajek1977}, the
question for asymptotic fields in magnetohydrodynamics was
posed. Requiring finite field and potential quantities at the horizon,
the so called Znajek `boundary condition' sets a link among the
angular derivative of $\Psi$ and the potential functions $I(\Psi)$ and
$\omega(\Psi)$ at the outer BH horizon:
\begin{align}
I_Z\left(\Psi\right)=-\frac{Mr_+\sin\theta}{r_+^2+a^2\cos^2\theta}\left[\Omega-\omega\left(\Psi\right)\right]\Psi_{,\theta}
\label{eq:ZnajekCondition}
\end{align} 
Despite its original purpose as a boundary condition, recent studies
suggest that eq.\,(\ref{eq:ZnajekCondition}) is a regularity condition
which is automatically satisfied in numerical procedures demanding
smoothness at the LS \citep{Komissarov2004,Nathanail2014}. Observing
the behavior of the $L^\infty$ norm of the difference between the
function $I\left(\Psi\right)$ and $I_Z(\Psi)$ for different stencils
at $r=r_+$, we are able to confirm the status of
eq.\,(\ref{eq:ZnajekCondition}) as a regularity condition, which is
automatically fulfilled throughout the numerically iterative procedure
with the imposed regularity at the LS (see
fig.\,\ref{fig:ConvergenceZ}). As we can see from that figure, the
reduction of the error in the preservation of the Znajek condition
happens for all the matching strategies presented in
sec.\,\ref{sec:Matching}. However, the error level in the preservation
of such condition is some orders of magnitude smaller when employing a
biased discretization of the second order radial derivatives,
regardless of the application of any smoothing procedure for $\Psi$ at
the light surfaces.

The initial error depends on the chosen spin factor $a$ and becomes
greater for BHs which are close to maximally rotating. The deviations
between the numerical solution and the Znajek condition are dominated
by the matching point between the BH horizon and the equator as well
as close to the axis of rotation, where an approximation of $\Psi$
becomes necessary \citep[cf.][]{Contopoulos2013}.

\section{Numerical Results}
\label{sec:NumericalResults}
\begin{figure}
	\centering
\includegraphics[width=0.49\textwidth]{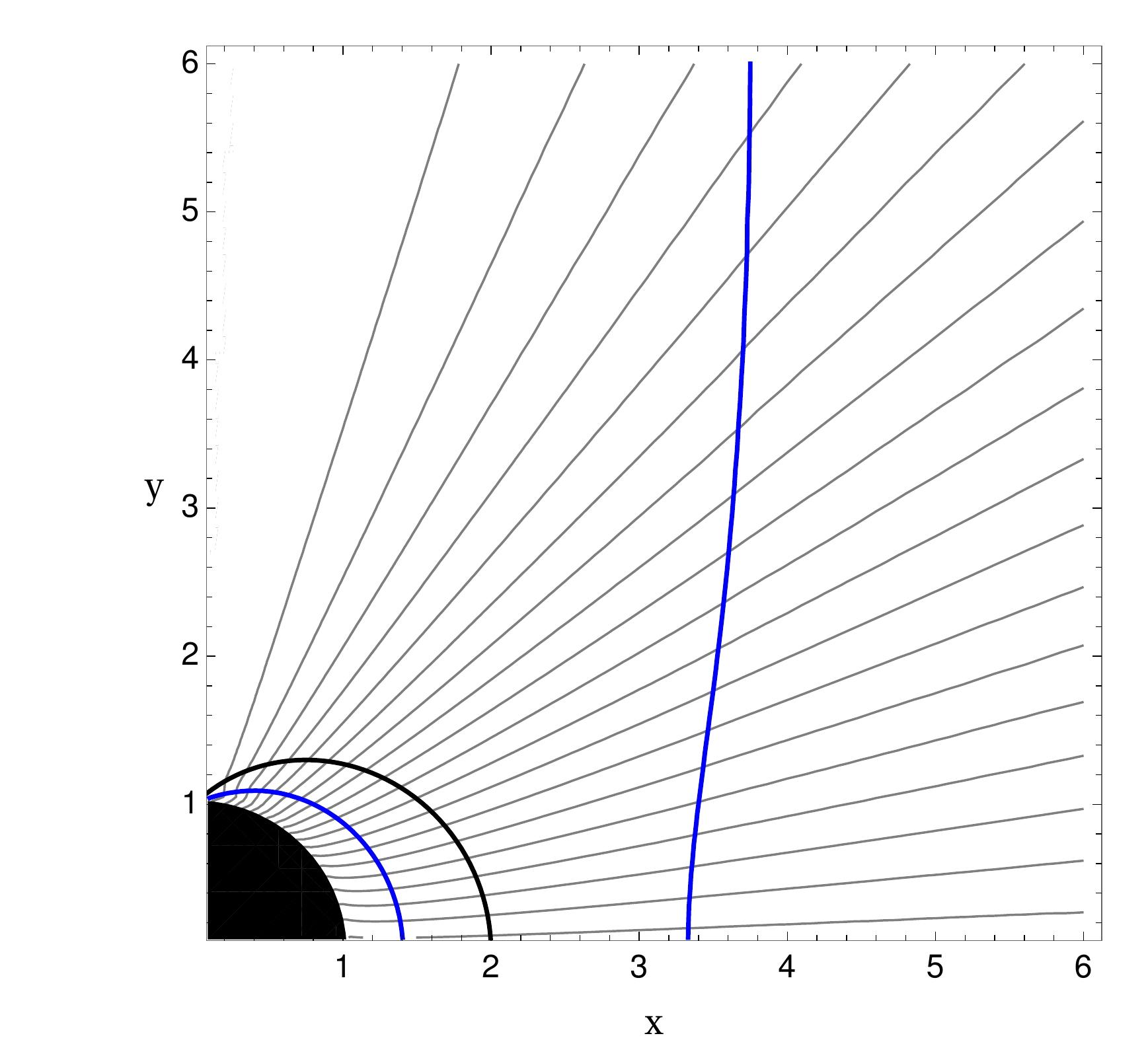} 
\caption{Distribution of the magnetic flux $\Psi$ in the vicinity of a
fast rotating ($a_*=0.9999$) BH. In order to reach the
configuration displayed, the GSE has been solved numerically until
the convergence criterion (\ref{eq:ConvCriterion}) has been reached
in a physical domain $[r_+,\infty] \times [0,90^\circ]$, covered
with a numerical grid
$\left[n_r\times n_\theta\right]=\left[200\times 64\right]$. The
location of the ergosphere is represented by the black line, the two
LS are drawn as blue lines. Magnetic flux configurations have been
studied for various spin parameters, some of which are visualized in
appendix \ref{sec:convergence}.}
\label{fig:SMParab}
\end{figure}

\subsection{Split monopole configurations}
\label{sec:smconfigs}
\begin{figure}
	\centering
%
	\includegraphics[width=0.42\textwidth]{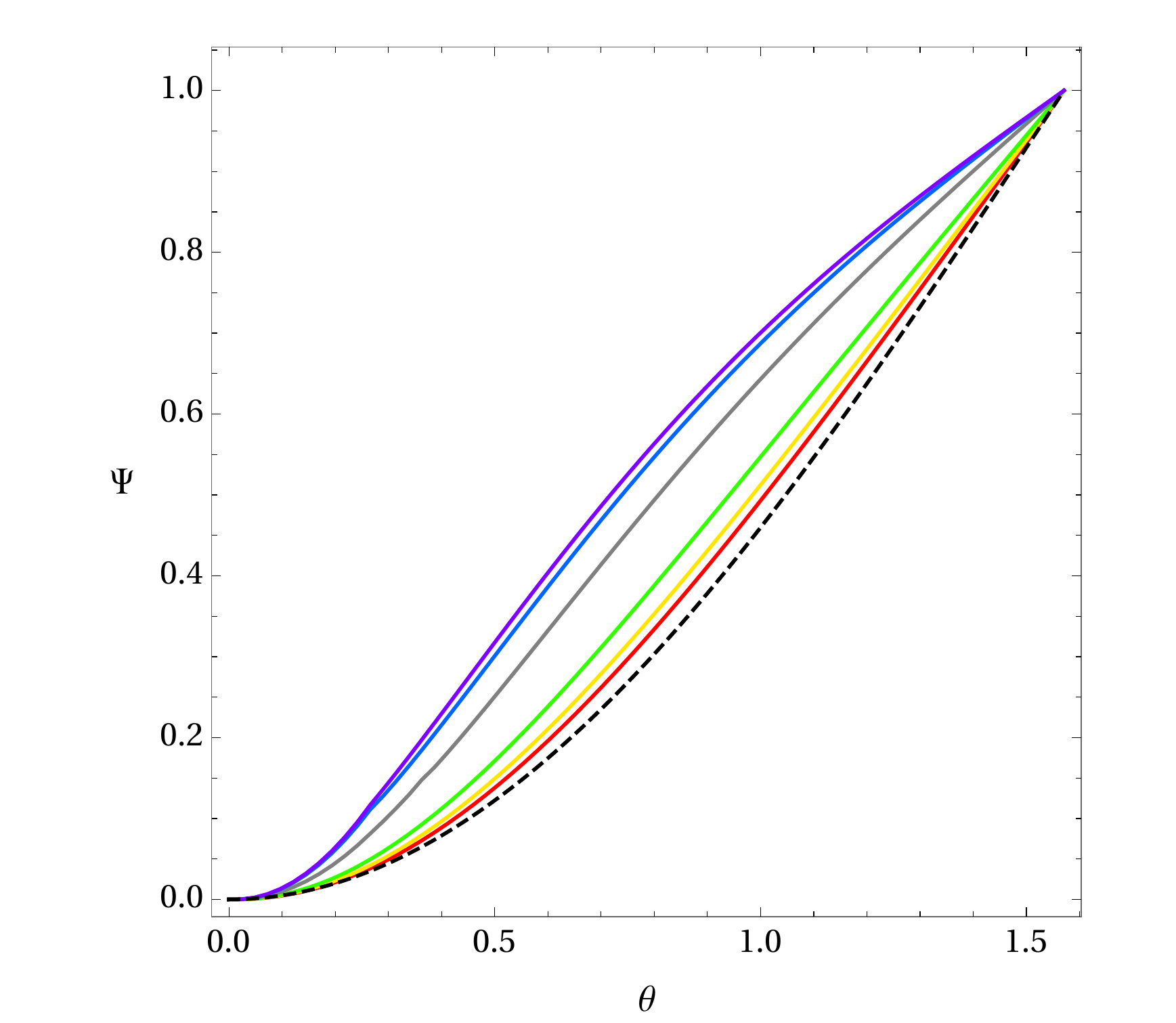} \\
	\vspace{5pt}
	\includegraphics[width=0.42\textwidth]{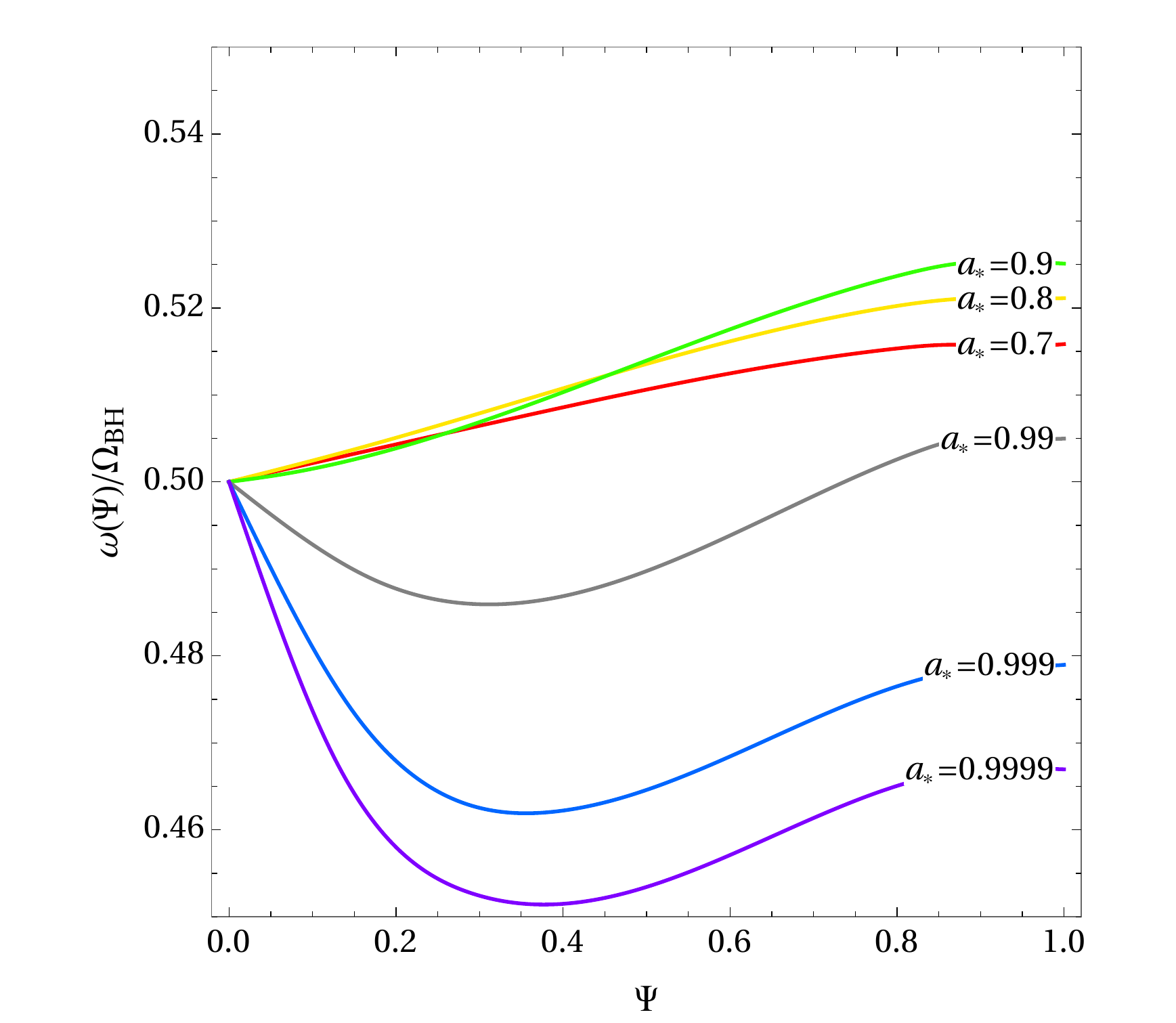} \\
	\vspace{5pt}
	\includegraphics[width=0.42\textwidth]{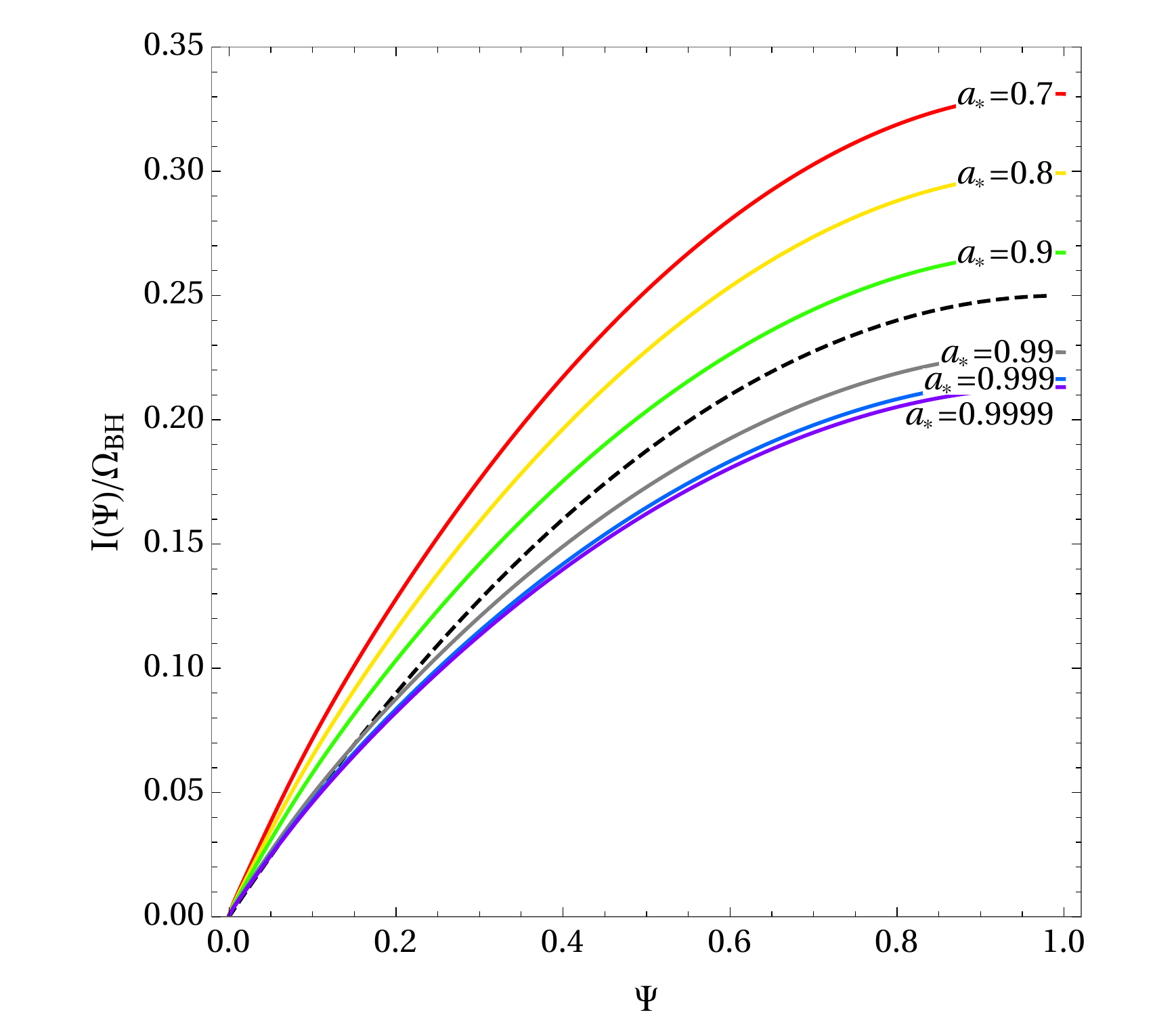} 
	\caption{Comparison of split-monopole solutions to the GSE for
 different spin parameters
 $a_*=\left\{0.7,0.8,0.9,0.99,0.999,0.9999\right\}$
 in a physical domain $[r_+,\infty] \times [0,90^\circ]$,
 covered with a numerical grid
 $\left[n_r\times n_\theta\right]=\left[200\times
 64\right]$. \textit{Top:} Angular distribution of the
 magnetic flux at the location of the inner light
 surface. The dashed line represents the initial values of
 the potential $\Psi_0$. \textit{Middle:} Distribution of
 $\omega\left(\Psi\right)$ after
 convergence. \textit{Bottom:} Distribution of
 $I\left(\Psi\right)$ after convergence.}
	\label{fig:SMQuantities}
\end{figure}
The first test for the numerical solution of the GSE is the
split-monopole \citep[cf.][]{Ghosh2000}, which has also been discussed
by many authors (e.g.,
\citealt{Komissarov2004,Contopoulos2013,Nathanail2014}). In
 the limit of a slowly rotating black hole, the split-monopole
 matches the flat spacetime solution of \cite{Michel1973} at large
 radii, while at the same time it satisfies the so called Znajek condition
 (eq.\,\ref{eq:ZnajekCondition}) at the event horizon. Admittedly,
 this solution is unphysical since in astrophysical conditions the
 magnetic field threading the horizon of a BH is supported by the
 electric currents in an accretion disc
 \citep[c.f.][]{Blandford1977,Komissarov2004}. Nevertheless, it is
 likely the simplest configuration that allows one to demonstrate the
 extraction of energy through the BZ
 mechanism\footnote{\cite{Komissarov2001} showed the action of the BZ
 mechanism for the first time in time-dependent, force-free
 numerical simulations on a static spacetime.}. The initial guess for the potential
$\Psi$ corresponds to a homogeneous solution of
eq.\,(\ref{eq:GSLightCylinder}) in the case of $a=0$ (also called the
Schwarzschild monopole, e.g.,\,\citealt{Ghosh2000}):
\begin{align}
\Psi_0\left(r,\theta\right)=1-\cos\theta,
\label{eq:GuessPsi}
\end{align}
where the maximum value of $\Psi$ has been normalized to 1.
In order to set the initial functional dependence of $\omega(\Psi)$
and $I(\Psi)$, we adopt the field line angular velocity as being half
the BH angular velocity $\omega=\OmegaBH/2$
\citep{Blandford1977}. For the currents we employ the analytical
solution of the pulsar magnetosphere. More specifically we set:
\begin{align*}
\omega_0\left(\Psi\right)=& \frac{1}{2}\frac{a}{r_+^2+a^2}\\
I_0\left(\Psi\right)= & -\frac{1}{\mbox{\addnew{2}}}\omega\left(\Psi\right)\Psi\left(2-\Psi\right)
\end{align*}
%
%
for $\Psi_{\rm min}\leq\Psi\leq \Psi_{\rm max}$, where
$\Psi_{\rm min}:=0$ and $\Psi_{\rm max}:=1$ are given by the potential
on the (Dirichlet) boundaries at the axis of rotation ($\theta=0$) and
the equator ($\theta=\upi/2$), respectively
(eq.\,\ref{eq:GuessPsi}). As stated before, Newman boundary conditions
are set up along the radial edges of the computational domain (i.e.,
at $r=r_+$ and at $r=r_{\rm max}$). 
Using eq.\,(\ref{eq:PotFieldConversion}) the
 corresponding initial magnetic fields are
\begin{align}
B^r(r,\theta)=&\: \frac{\sin\theta}{2\upi\sqrt{\gamma}},\\
B^\theta(r,\theta)=&\: 0,\\
B^\phi(r,\theta)=&\: \frac{a}{4(r_+^2+a^2)}\sqrt{\frac{\Sigma}{A\Delta}}.
\label{eq:GuessB}
\end{align}
Figure~\ref{fig:SMParab} shows
the topology of the magnetic flux $\Psi$ computed solving numerically
the GSE until our convergence criterion (eq.\,\ref{eq:ConvCriterion})
is reached. We display the case of a nearly maximally rotating BH
($a_*=0.9999$) in fig.\,\ref{fig:SMParab}. The isocontours of $\Psi$
pass smoothly through both LS (displayed with thick blue lines). A
comparison of solutions to the GSE for different spin parameters is
shown in fig.\,\ref{fig:SMQuantities}.
 
 In all tests where $\omega$ and $I$ are let to relax from their
 initial values, the error measures $\mathcal{R}_\Psi$ and
 $\mathcal{R}_{\rm LC}$ can be reduced substantially after a
 sufficiently large number of iterative steps. This demonstrates that
 our method is robust and converges to a smooth, numerically stable
 solution. For this to happen, a number of technical comments are
 crucial at this point. First, we find convergence to a smooth
 solution if we set a double convergence criterion both on
 $\mathcal{R}_\Psi$ and $\mathcal{R}_{\rm LC}$ (see
 app.\,\ref{sec:convergence}). In contrast to \cite{Nathanail2014}, if
 we restrict our solution procedure to relax the initial set up for
 3000-4000 iterations, one does not find a steady state solution. As
 shown in the bottom panels of fig.\,1 of \cite{Nathanail2014} the
 solution displays kinks close to the ILS. Many more iterations are
 necessary to qualify the numerical solution as a `steady state' (for
 details, see app.\,\ref{sec:technical_notes}).

\subsection{Paraboloidal configurations}
\label{sec:Paraboloidal}
\begin{figure}
	\centering
	\includegraphics[width=0.49\textwidth]{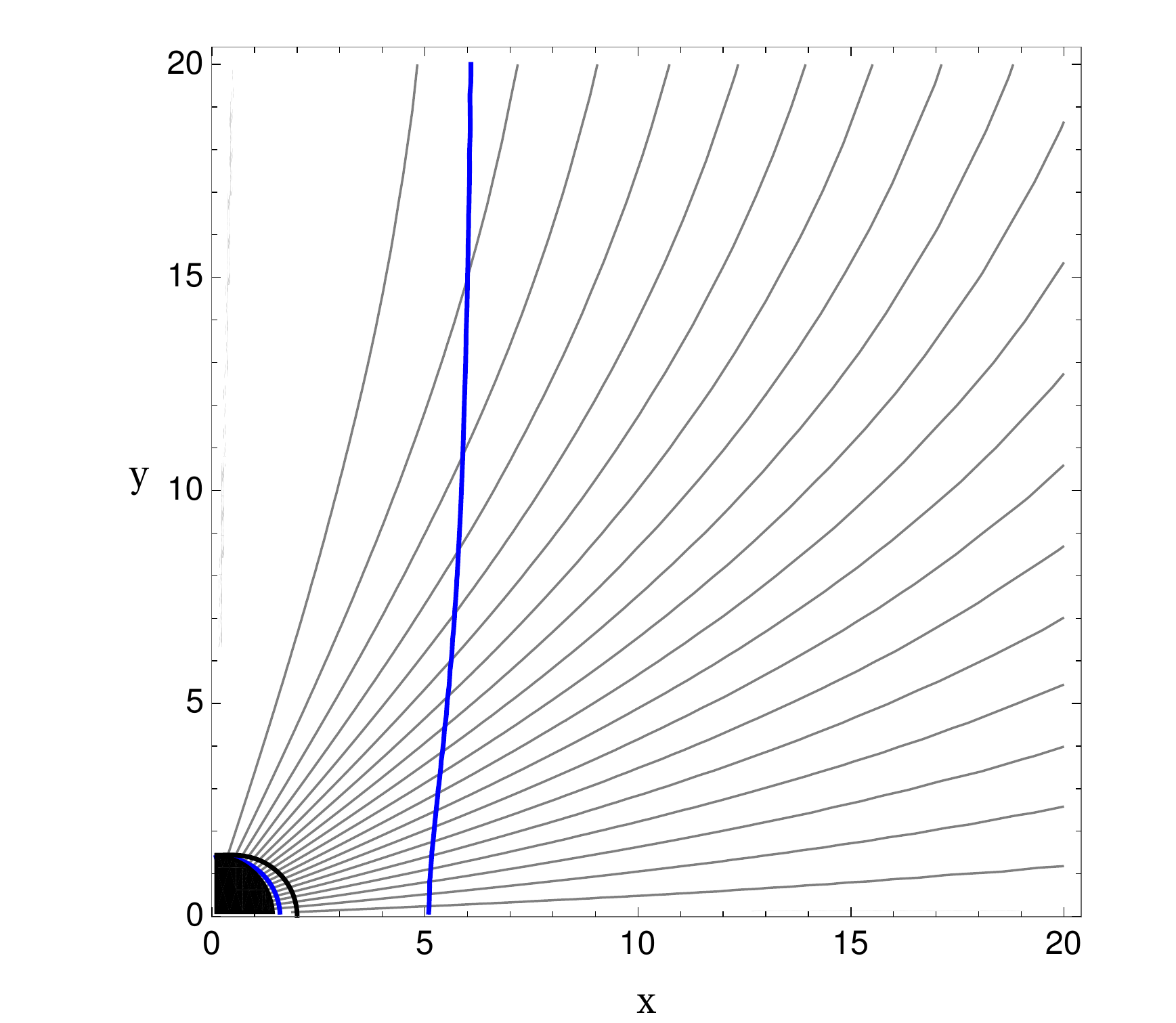}
	\includegraphics[width=0.49\textwidth]{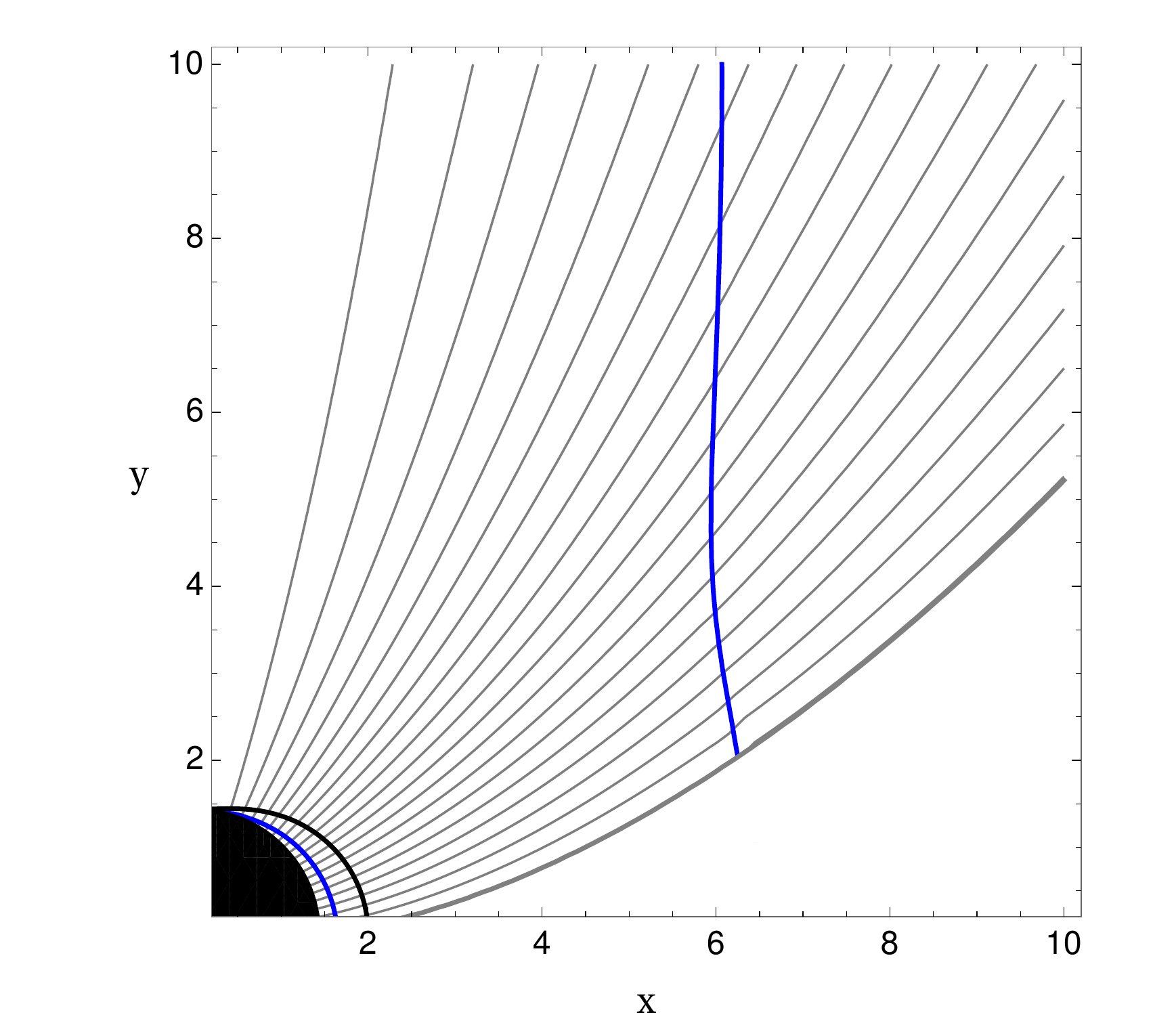}
	\caption{Distribution of the magnetic flux $\Psi$ in the
 vicinity of a fast rotating $a_*=0.9$ BH employing
 appropriate boundary conditions for paraboloidal confinement
 (cf. sec. \ref{sec:Paraboloidal}) in a physical domain
 $[r_+,100] \times [0,90^\circ]$, covered with a numerical
 grid
 $\left[n_r\times n_\theta\right]=\left[200\times
 64\right]$. \textit{Top:} Far boundary according to
 \citet{Fendt:1997A&A...319.1025}. \textit{Bottom:}
 Paraboloidal confinement following the \citet{Nathanail2014}
 setup ($r_0=10$).}
	\label{fig:Parab}
\end{figure}
\begin{figure}
	\centering
	\includegraphics[width=0.49\textwidth]{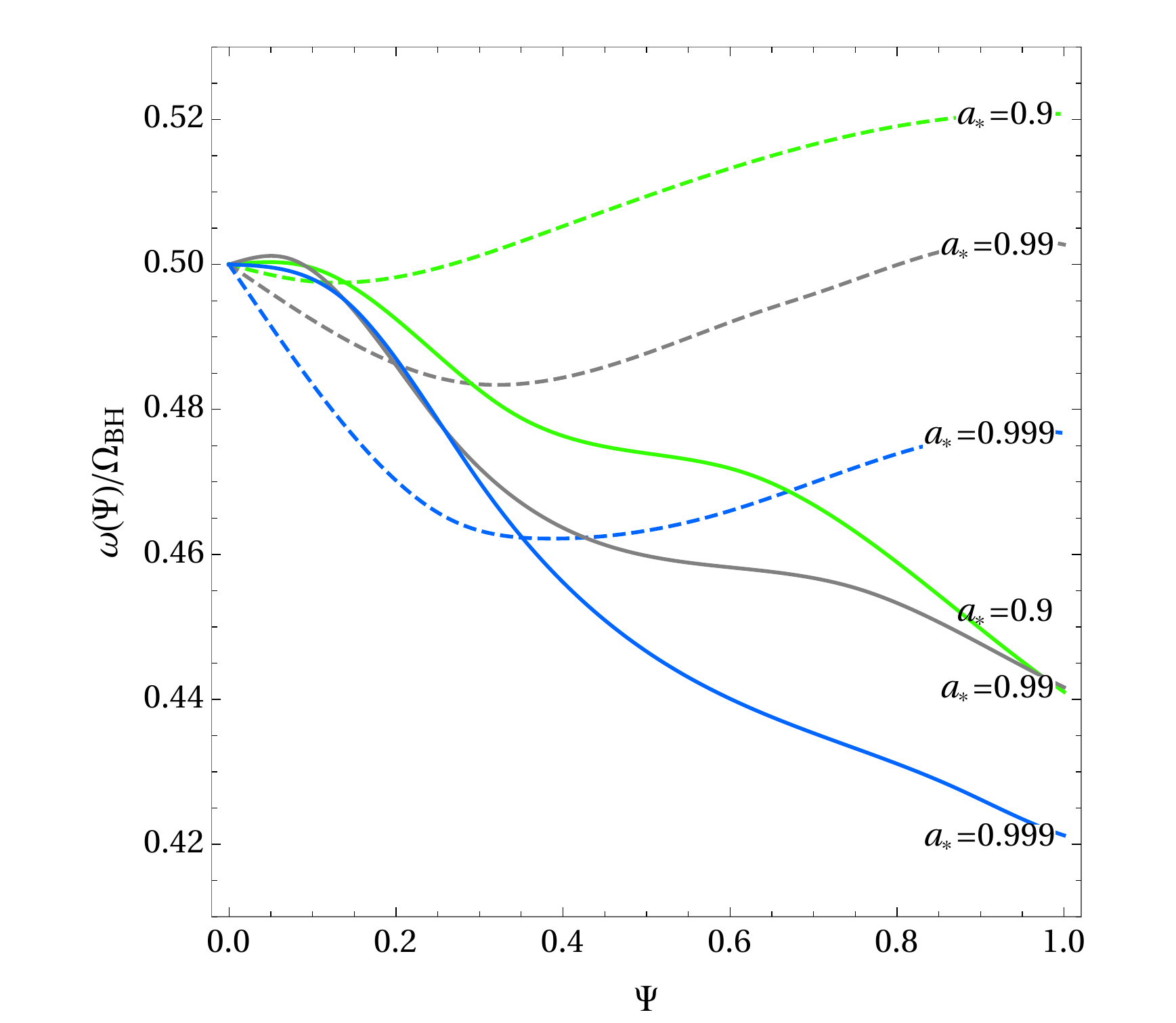}
	\includegraphics[width=0.49\textwidth]{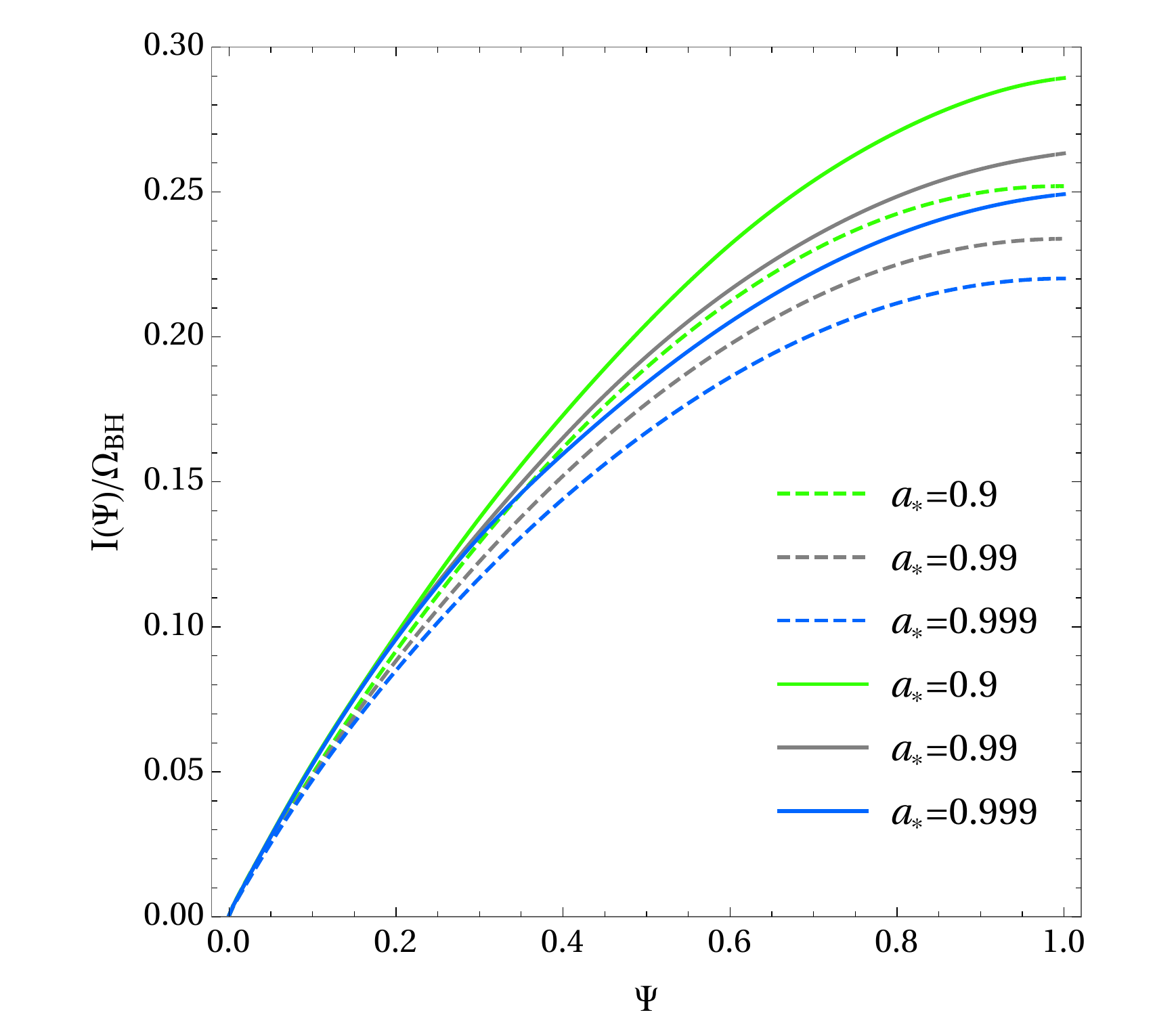}
	\caption{\addnew{Final distributions (after convergence is
 reached) of the scalar functions $\omega\left(\Psi\right)$
 (top panel) and $I\left(\Psi\right)$ (bottom panel) in the
 vicinity of fast rotating BHs of different spin parameters
 $a_*=\left\{0.9,0.99,0.999\right\}$ embedded in
 paraboloidal magnetic field magnetosphere
 (cf. sec. \ref{sec:Paraboloidal}). The physical domain
 $[r_+,100] \times [0,90^\circ]$ is covered with a
 numerical grid
 $\left[n_r\times n_\theta\right]=\left[200\times
 64\right]$. The relaxed solution of the
 \citet{Fendt:1997A&A...319.1025} approach is represented
 with dashed lines, the \citet{Nathanail2014} strategy by
 solid lines.} }
	\label{fig:ParabOmegaI}
\end{figure}
Collimated magnetospheres are an important ingredient of jet outflows
from compact objects and may be found, e.g., in the parabolically
shaped solutions to the GSE which have been studied by, e.g.,
\cite{Fendt:1997A&A...319.1025} and
\cite{Nathanail2014}. \cite{Fendt:1997A&A...319.1025} considers an
asymptotically cylindrical shape of the magnetic field and suggests to
set up the outer radial boundary at $r_{\rm max}$ according to
\begin{align}
 \Psi_{\rm out}(x)=\frac{1}{b}\ln\left(1+\frac{x^2}{d^2}\right),
\label{eq:ParabBound}
\end{align}
where the constants $d$ and $b$ determine the degree of collimation
and $x=r\times\sin\theta$. \cite{Fendt:1997A&A...319.1025} adapts his
finite element computational domain to provide a parabolically shaped
outer jet boundary. Indeed, we have considered setting up Fendt's
boundaries at radial infinity, resulting into a final solution which
resembles that of a split monopole (i.e., effectively
unconfined). Bringing condition (\ref{eq:ParabBound}) to a finite
distance, solutions with a degree of confinement larger than the split
monopole case are possible. For instance, in fig.\,\ref{fig:Parab}
(upper panel) we show the solution for $\Psi$ when setting the
Dirichlet boundary condition (\ref{eq:ParabBound}) at
$r_{\rm max}=100$ for values of the confinement parameters $b=9.2$,
$d=1.0$.

The numerical relaxation proceeds without obstacles and ultimately
yields solutions, which show higher confinement at the OLS (comparing
to the split-monopole solutions in sec.\,\ref{sec:smconfigs}). It
should be noted, that an appropriate choice of boundary conditions at
$r_{\rm max}$ is crucial for the relaxation towards a paraboloidal
setup. Using a parabolic topology as initial guess but with Newman
boundary conditions (zero derivative) at $r_{\rm max}$ and no further
induced confinement will result in a split-monopole solution.

However, setting Fendt's condition at finite distance is artificial
and difficult to justify in astrophysical BH magnetospheres. As an
alternative, \citet{Nathanail2014}%
\footnote{\addnew{See also \cite{Tchekhovskoy2010} for an equivalent
 set up in FFDE and GRMHD.}}
suggest to solve the GSE for a confined parabolic setup limiting the
computational domain to a region $0\leq\Theta\leq 1$, where
\begin{align*}
\Theta\left(r,\theta\right)=\frac{\theta}{\theta_{\rm wall}\left(r\right)},
\end{align*}
and $\theta_{\rm wall}\left(r\right)$ describes a paraboloidal wall according to the function
\begin{align}
1-\cos\theta_{\rm wall}=\left(\frac{r+r_0}{r_+ + r_0}\right)^{-\nu}.
\label{eq:ParabWall}
\end{align}
Here, $r_0$ and $\nu$ are parameters determining the degree of
confinement of the parabolic boundary wall, e.g., a choice of
$r_0=\infty$ or $\nu=0$ reduces to the split monopole initial guess in
eq.\,(\ref{eq:GuessPsi}). Employing $\Theta\left(r,\theta\right)$ to
define the angular coordinate \citep[c.f.,][]{Nathanail2014}, allows
us to use the numerical solver as in previous examples without the
need for {\em excising} regions of the computational domain in the
vicinity of the equatorial plane in order to ensure the paraboloidal character of the solution. Employing the function
\begin{align*}
\theta\left(r,\Theta\right)=\Theta\times\theta_{\rm wall}\left(r\right),
\end{align*}
where $\Theta\in\left[0,1\right]$, as well as the following coordinate changes in the angular derivatives
(cf. eq.\,(\ref{eq:CoordChangeR}) for the radial derivatives)
\begin{align*}
\begin{split}
\frac{\partial}{\partial\theta}=&\:\frac{1}{\theta_{\rm wall}\left(r\right)}\frac{\partial}{\partial\Theta}\\
\frac{\partial^2}{\partial\theta^2}=& \:\left(\frac{1}{\theta_{\rm wall}\left(r\right)}\right)^2\frac{\partial^2}{\partial\Theta^2},
\end{split}
\end{align*}
no further changes to the update and LS routines are necessary. After setting up an initial guess for the potential according to
\citep[following][]{Nathanail2014}
\begin{align}
\Psi\left(r,\theta\right)=\Psi_{\rm max}\left(\frac{r+r_0}{r_+ + r_0}\right)^\nu\left(1-\cos\theta\right), 
\label{eq:Psiparaboloid2}
\end{align}
\addnew{with $\Psi_{\rm max}=1$ for simplicity,} as well as
$\omega\left(\Psi\right)$ and $II'\left(\Psi\right)$ like in
sec.\,\ref{sec:smconfigs}, the numerical solution converges without
obstacles. We observe, however, that with larger confinement (i.e.,
lower values of $r_0$), the kinks at the OLS become stronger,
especially close to the equatorial plane. The growing artifacts may be
reduced with lower relaxation factors in the SOR routines. The
presented example (fig.\,\ref{fig:Parab} bottom panel) shows the
converged solution (demanding condition \ref{eq:ConvCriterion}) for
$\omega_{\rm SOR}=1.0$ after $42.740$ iterations. \addnew{The solid
 lines in fig.\,\ref{fig:ParabOmegaI} display the distributions of
 $\omega\left(\Psi\right)$ and $I\left(\Psi\right)$ after convergence
 is reached employing the set up of \cite{Nathanail2014}. For
 comparison, we also display (with dashed lines) the final
 distributions of $\omega\left(\Psi\right)$ and $I\left(\Psi\right)$
 employing the paraboloidal problem set up of
 \cite{Fendt:1997A&A...319.1025}. From the top panel of
 fig.\,\ref{fig:ParabOmegaI}, it is evident that the latter set up
 tends to produce solutions with faster rotating field lines, not
 only in the equatorial plane (corresponding to a value $\Psi=1$),
 but at almost every other latitude in the domain ($\Psi=0$
 corresponds to the axis of symmetry).}

\newpage
\subsection{Vertical field configurations}
\label{sec:VertField}
\begin{figure*}
	\centering
	\includegraphics[width=1.0\textwidth]{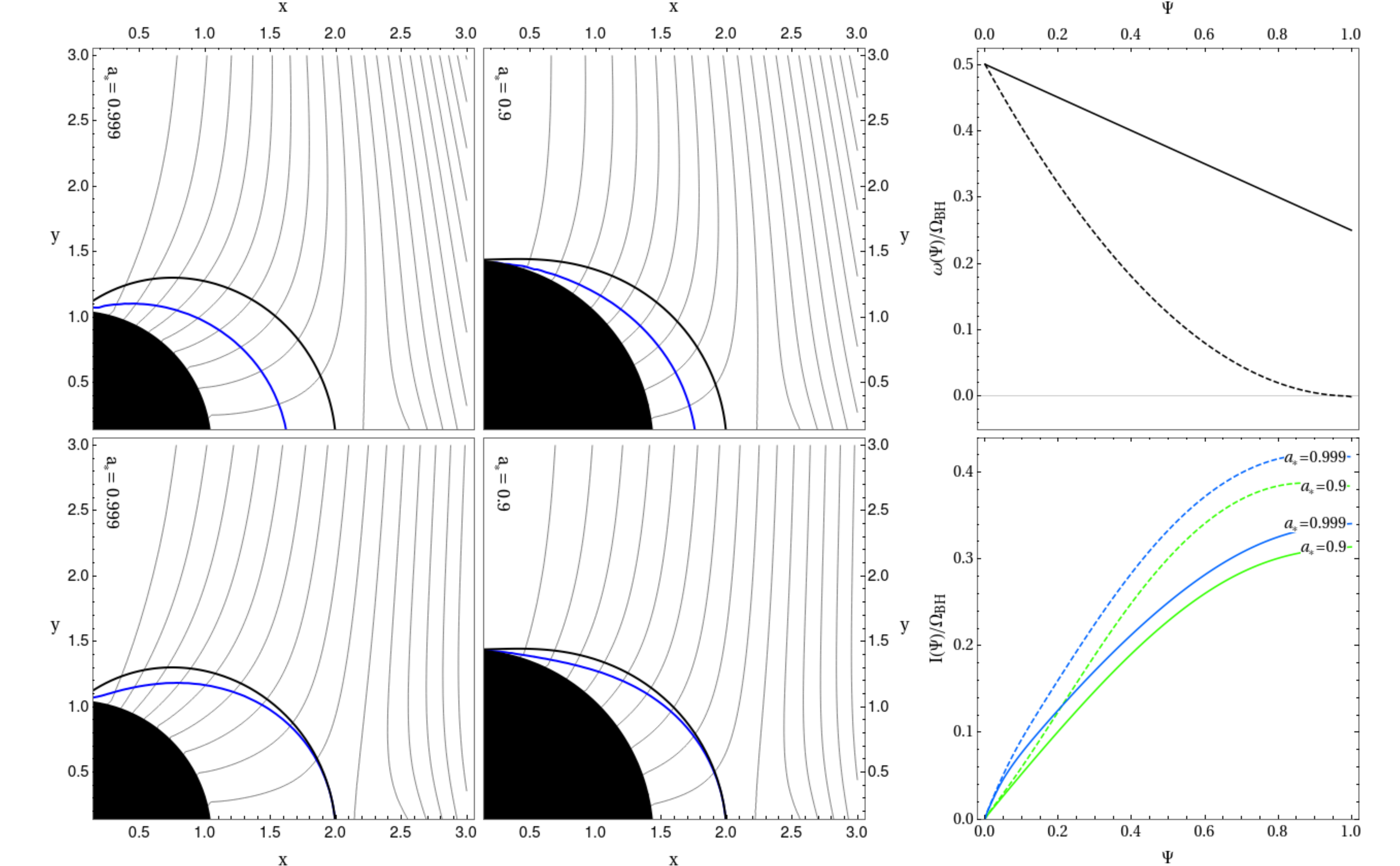}
	\caption{\addnew{Distribution of $\Psi$ (left and central
 columns), of $\omega\left(\Psi\right)$ (top right panel)
 and of $I\left(\Psi\right)$ (bottom right panel) in the
 vicinity of fast rotating BHs of different spin parameters
 $a_*=\left\{0.9,0.999\right\}$ embedded in a vertical
 magnetic field (cf. sec. \ref{sec:VertField}). The
 physical domain $[r_+,100] \times [0,90^\circ]$ is covered
 with a numerical grid
 $\left[n_r\times n_\theta\right]=\left[200\times
 200\right]$. In the top right panel we depict the
 (fixed) fieldline profiles of $\omega\left(\Psi\right)$
 given by eq. (\ref{eq:VertOm2}) (\textit{solid line}) and
 eq. (\ref{eq:VertOm1}) (\textit{dashed line}). The upper
 left and middle panels show the spatial distribution of
 $\Psi$ corresponding to the imposed profile of
 $\omega\left(\Psi\right)$ given by eq. (\ref{eq:VertOm2}).
 The bottom left and middle panels show the spatial
 distribution of $\Psi$ corresponding to
 $\omega\left(\Psi\right)$ given by eq. (\ref{eq:VertOm1}).
 The relaxed values of $I\left(\Psi\right)$ corresponding
 to the two (fixed) distributions of
 $\omega\left(\Psi\right)$ are shown on the bottom right
 panel (using the same line styles as in the upper right
 panel).}}
	\label{fig:VertFields}
\end{figure*}
\addnew{
	
 Another well studied exemplary fieldline configuration is the
 embedding of a BH into a vertical magnetic field. Originally
 considered by \citet{Wald1974} for the electrovacuum limit, it has
 since been studied in dynamical evolutions \citep[see,
 e.g.,][]{Komissarov2004,Komissarov2007a,Palenzuela2010} as well as
 with a focus on the BH "Meissner effect"
 \citep{Komissarov2007a,Nathanail2014}, or on the uniqueness of the
 numerical solutions \citep{Pan:2017}. The case of vertical
 fieldlines opens up the possibility of fieldlines crossing only the
 ILS \citep[for a detailed discussion, see also][]{Nathanail2014}
 and, hence, the freedom of fixing either $\omega\left(\Psi\right)$
 or $I\left(\Psi\right)$ throughout the relaxation procedure
 (cf. sec. \ref{sec:UpdatesPotentials}).
	
 We employ a Dirichlet boundary at $\theta=0$, as we have done in
 sec.\,\ref{sec:smconfigs}. In order to fill up the initial magnetic
 field configuration we divide the computational domain into three
 regions. The first region is the spherical shell surrounding the BH and
 extending slightly beyond the ergosphere defined by
\begin{align*}
r < 1.25\times r^*_+(\upi/2) \equiv r_0\:,
\end{align*}
where we employ the split monopole potential $\Psi_0$
(eq.\,\ref{eq:GuessPsi}). The second region extends beyond the
previous spherical shell up to infinity in the vertical direction,
i.e. enclosing the region
\begin{align*}
r_0\le r < r\times\sin\theta \:,
\end{align*}
where the isolines of $\Psi$ are vertical and their values correspond
to
\begin{align*}
\Psi(r,\theta) = \Psi_0\left(r, \arcsin\left( \frac{r\times \sin\theta}{r_0} \right)\right)\:.
\end{align*}
Finally, in the third region, defined by $r\times\sin\theta > r_0$, we
use
\begin{align*}
\Psi\left(r,\theta\right)=\left(\frac{r\times\sin\theta}{r_0}\right)^2.
\end{align*}
We apply the following Dirichlet boundary condition at the equator:
\begin{align*}	
\Psi\left(r,\theta=\frac{\upi}{2}\right)=\left\{\begin{array}{cc}
1.0 & \text{if } r<r_0\\[1em]
\displaystyle\left(\frac{r}{r_0}\right)^2 & \text{otherwise}
\end{array}\right.
\end{align*}
Newman boundary conditions are set up along the radial edges of the
computational domain (i.e. at $r=r_+$ and at $r=r_{\rm
 max}$). Following \citet{Nathanail2014}, we initiate the fieldline
angular velocity to one of the two functions
\begin{align}
\omega\left(\Psi\right)&=\frac{\OmegaBH}{2}\times\left(1-\Psi\right)^2 \label{eq:VertOm1}\\
\omega\left(\Psi\right)&=\:\frac{\OmegaBH}{2}\times\left(1-\frac{\Psi}{2}\right) \label{eq:VertOm2},
\end{align}
for $\Psi_{\rm min}\leq\Psi\leq \Psi_{\rm max}$, where
$\Psi_{\rm min}:=0$ and $\Psi_{\rm max}:=1$, and zero otherwise. This
choice of $\omega\left(\Psi\right)$ pushes the OLS to infinity and
allows us to update only the current function $I\left(\Psi\right)$
throughout the numerical relaxation. By the construction of the
boundary conditions, eq. (\ref{eq:VertOm1}) ensures that the ILS
touches the outer ergosphere $r_+^*$ at the equator, while
eq. (\ref{eq:VertOm2}) provides an ILS well inside the outer
ergosphere $r_+^*$.

The initial values employed for the numerical algorithm and the
equatorial boundary conditions slightly differ from those employed in
\cite{Nathanail2014}, but they provide smooth profiles of the ILS and
no glitches in the field lines in the region enclosed in between of
the ILS and the outer ergosphere (see the left and middle panels of
fig.\,\ref{fig:VertFields}). These results can be compared with the
ones presented by \citeauthor{Nathanail2014}
(\citeyear{Nathanail2014}; fig.\,3) or by \citeauthor{Pan:2017}
(\citeyear{Pan:2017}; fig.\,1). The fact that the configurations we
find are slightly different to those of the previous works is simply a
consequence of the different boundary conditions employed, and not
necessarily related to a lack of uniqueness of the GSE solution
\citep[see][and Sec.\,\ref{sec:discussion} for a discussion of the
uniqueness of the GSE solution for the vertical magnetic field
configuration]{Pan:2017}.

We have further explored the influence of the equatorial boundary
conditions by setting up an approximately vertical magnetic field
employing a paraboloidal setup as discussed in
sec. \ref{sec:Paraboloidal}, e.g. by using the parameters $r_0=0$, and
$\nu=2$ in eqs. (\ref{eq:ParabWall}) and (\ref{eq:Psiparaboloid2}). The
relaxed solutions are as smooth as those obtained with the previous
initialization for $\Psi$ and the corresponding equatorial boundary
conditions. In none of the two cases the Meissner effect is observed,
in full agreement with the findings of e.g.\,\cite{Nathanail2014} or \cite{Pan:2016}.}

\subsection{BH-disk models}
\label{sec:BHdisk}
\begin{figure}
	\centering
	\includegraphics[width=0.49\textwidth]{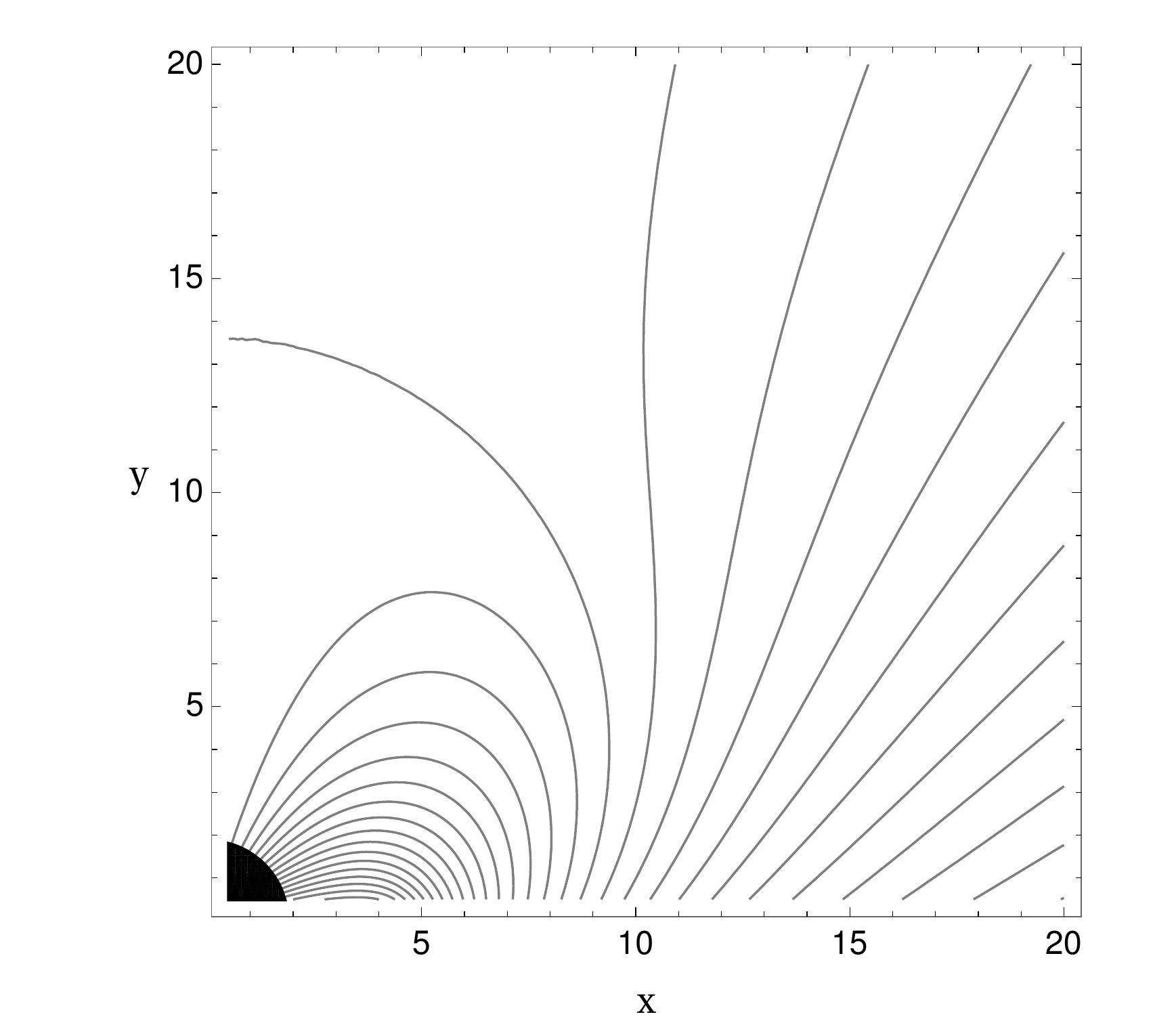}
	\includegraphics[width=0.49\textwidth]{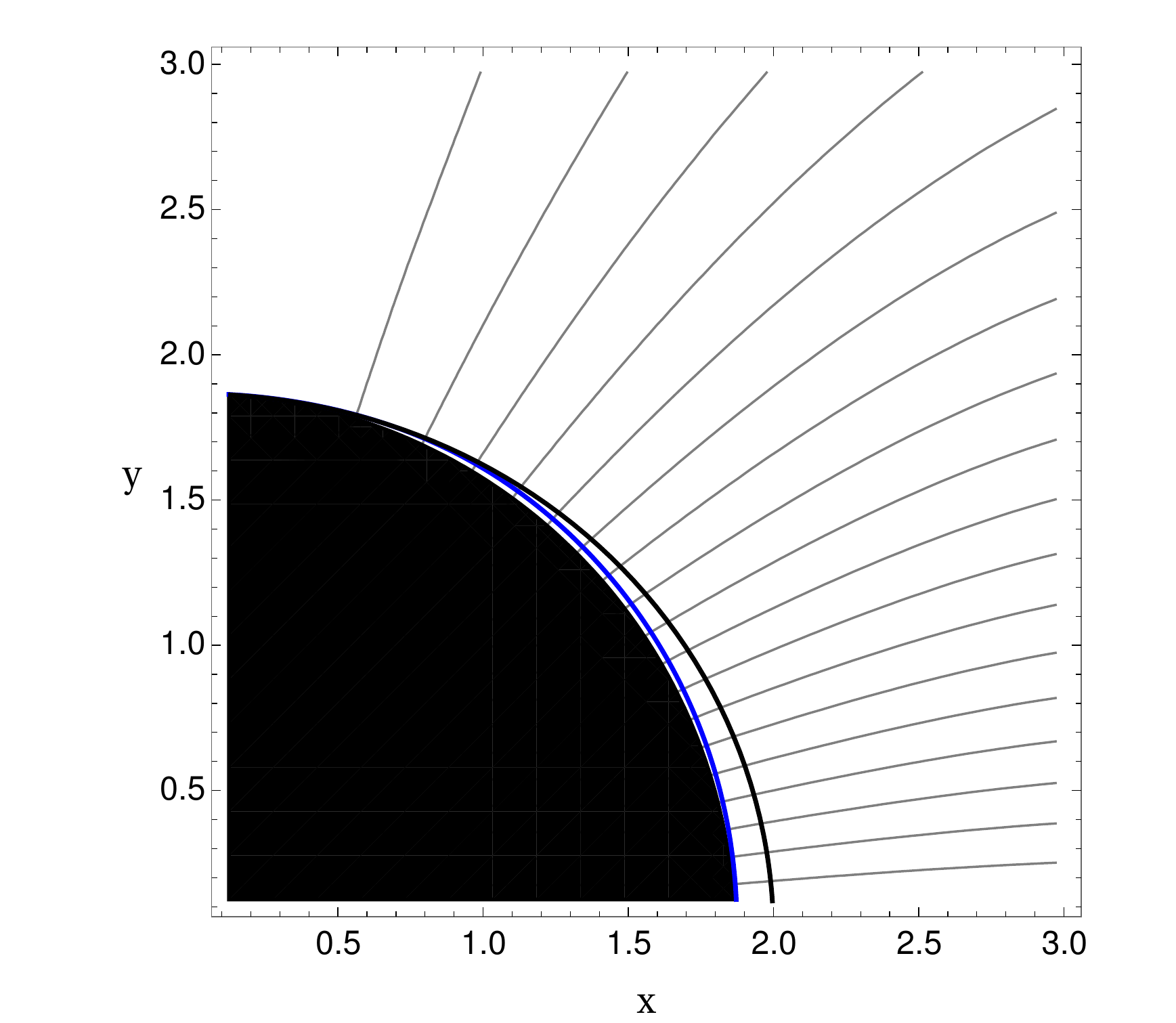}
	\includegraphics[width=0.49\textwidth]{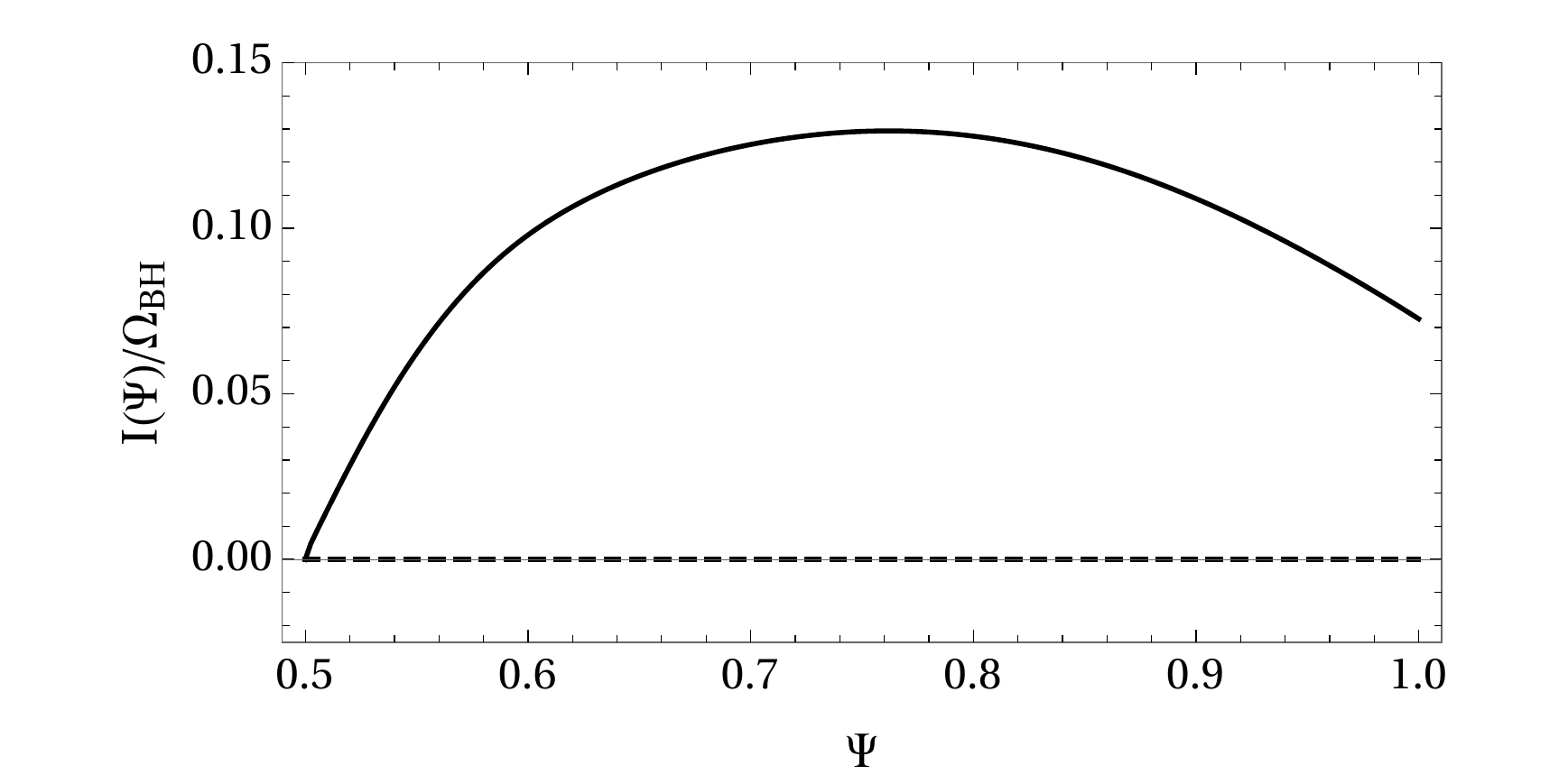}
	\caption{BH-disk model following the setup by
          \citet{Uzdensky:2005ApJ...620..889} after the relaxation
          procedure
          ($a_*=0.5, [r_+,\addnew{100}] \times
          [0,90^\circ],\left[n_r\times n_\theta\right]=\left[200\times
            200\right]$). The fieldlines connecting the BH to the disk
          rotate with Keplerian velocity. The open fieldlines are free
          of rotation and current. \addnew{\textit{Top/Middle:} Global
            structure and zoom of the inner region showing in detail
            the location of the ILS (blue line) in between of the
            ergosphere (thick black line) and the outer event
            horizon. \textit{Bottom:} Distribution of
            $I\left(\Psi\right)$ after convergence. The dashed line
            represents the initial configuration
            $I_0\left(\Psi\right)=0$.}}
	\label{fig:UZDParab}
\end{figure}
\cite{Uzdensky:2005ApJ...620..889} suggests the setup of a BH-disk
system via a suitable choice of boundary conditions in the numerical
solution of the GSE. In these configurations, field lines threading the
BH horizon connect to the equatorial plane and rotate with Keplerian
velocity. Following \cite{Uzdensky:2005ApJ...620..889}, the boundary
along the axis of rotation and the equatorial plane are set up as
follows:
\begin{align*}
\Psi\left(\theta=0\right)&=\Psi_{\rm s}\\
\Psi\left(r>r_{\text{in}},\theta=\frac{\upi}{2}\right)&=\Psi_{\rm d}\left(x\right)\\
\Psi\left(r\leq r_{\text{in}},\theta=\frac{\upi}{2}\right)&=\Psi_{\rm max}.
\end{align*}
Here, $r_{\text{in}}=r_{\text{ISCO}}\left(a\right)$ is the radial
location of the innermost circular orbit as a function of
$a$. $\Psi_{\rm s}$ fixes the value of the separatrix between open
field lines and field lines linking the BH to the disk. Their potential
is connected to the disk radius by
\begin{align*}
\Psi_{\rm d}\left(r\right)=\Psi_{\text{max}}\left(\frac{r_{\text{in}}}{r}\right).
\end{align*}
Homogeneous Neumann boundary conditions are set at the outer event
horizon,
\begin{align*}
\Psi_{,r}\left(r=r_+\right)=0.
\end{align*}
We find that it is necessary to fix the boundary at $r_{\rm max}$ to a
predefined shape function similar to the paraboloidal case presented
in sec.\,\ref{sec:Paraboloidal}:
\begin{align*}
\Psi\left(r=r_{\rm max}\right)=\Psi_{\rm s}+\left[\Psi_{\rm d}\left(r\right)-\Psi_{\rm s}\right]\left(1-\cos\theta\right).
\end{align*}
As for the field line angular velocity $\omega$, it is assumed that the
field lines connected to the BH rotate with the Keplerian angular
velocity $\Omega_{\rm K}$ of the disk, 
\begin{align*}
	\Omega_{\rm K}\left(r\right)=\frac{\sqrt{M}}{r^{3/2}+a\sqrt{M}},
\end{align*}
and do not rotate for
$\Psi<\Psi_{\rm s}$, namely,
\begin{align*}
\omega\left(\Psi\right)=\left\{\begin{array}{cc}
0 & \text{if } \Psi<\Psi_{\rm s}\\[1em]
\Omega_{\rm K}\left[r_0\left(\Psi\right)\right]
 \displaystyle\tanh^2\left(5\frac{\Psi-\Psi_{\rm
s}}{\Psi_{\text{tot}}-\Psi_{\rm s}}\right) & \text{otherwise}
\end{array} \right.
\end{align*}
where $r_0$ represents the footpoint of a given field line with
potential $\Psi$ on the disk.

Following \citet{Uzdensky:2005ApJ...620..889}, we use a grid of
$\left[n_r\times n_\theta\right]=\left[200\times 200\right]$ numerical
nodes. However, relatively small values of $a$ let the ILS approach
the outer BH event horizon $r_+$, rendering insufficient the previous
radial grid resolution \addnew{(see the LS position in the mid panel
 of fig. \ref{fig:UZDParab})}. Therefore, we choose to \addnew{adapt
 the radial coordinate for the discretization to the function $R_{\rm min}=r_{+}/\left(r_{+}+10M\right)$, and} refine the grid
increasing by a factor of 20 the number of nodes in the radial
direction whenever we update the potential functions $I\left(\Psi\right)$ and $\omega\left(\Psi\right)$, e.g. we employ
$\left[n_r\times n_\theta\right]=\left[4000\times 200\right]$ for
$a_*=0.5$ in order to ensure sufficient data points around the ILS. As
$I\left(\Psi\right)=0$ and $\omega\left(\Psi\right)=0$ for
$\Psi<\Psi_{\rm s}$, the OLS is pushed to infinity and the
magnetosphere is current-free along the open field lines. The numerical
relaxation proceeds smoothly towards a converged equilibrium solution
(see fig.\,\ref{fig:UZDParab}), which closely matches that found by
\cite{Uzdensky:2005ApJ...620..889}.

\section{Power of the BZ process}
\label{sec:BZpower} 
%
\begin{figure}
	\centering
	\includegraphics[width=0.49\textwidth]{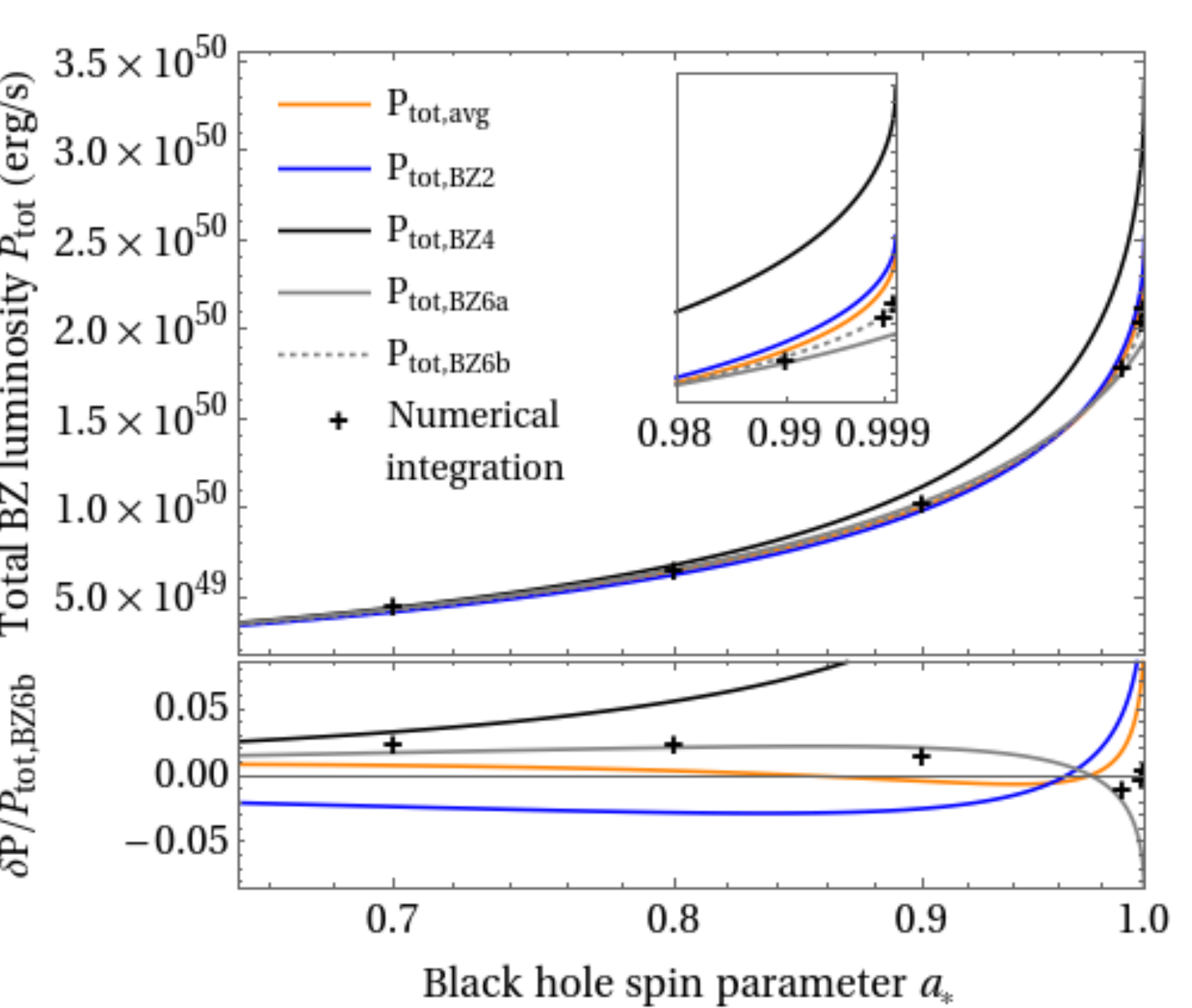}
	\caption{Comparison of the power of split-monopole solutions
 to the GSE in a physical domain
 $[r_+,\infty] \times [0,90^\circ]$, covered with a numerical
 grid
 $\left[n_r\times n_\theta\right]=\left[200\times
 64\right]$. \addnew{The figure shows} the total process
 power derived for different BH spin parameters of a
 $1 M_\odot$ BH computed using the direct numerical
 evaluation of eq.\,(\ref{eq:Ptot}) (black crosses), the
 approximated value of \citeauthor{Lee2000}
 (eq.\,\ref{eq:PtotLee}; orange line)\addnew{, as well as the
 second order (blue line) and 6th order (gray line)
 approximations of the BZ power as suggested by
 \citet{Tchekhovskoy2010}. The 6th order approximation
 (eq. \ref{eq:BZ6}) fitted to the numerical data (following
 eq. \ref{eq:UnitConversion}) is denoted by the dotted
 line. The extremal spin segment is magnified within the
 plot.}}
	\label{fig:Luminosity}
\end{figure}
\begin{table}
	\centering
\addnew{
	\begin{tabular}{c|c|c|c|c}
	Approximation & $P_{\rm tot, avg}$ &$P_{\rm tot, BZ2}$&$P_{\rm tot, BZ4}$& $P_{\rm tot, BZ6a}$ \\\hline
RMS error & 9.23 & 13.21 & 50.49& 7.19\\
		\hline
		\hline
	\end{tabular}
}
	\caption{\addnew{RMS error of the different
 approximations computed for the total BZ power presented
 in fig.\,\ref{fig:Luminosity}. The results are normalized
 to the RMS of the 6th order approximation computed using
 eq.\,(\ref{eq:BZ6}) and the fit parameters $b_1=0.81$ and $b_2=-5.61$ (BZ6b).}}
	\label{tab:rms}
\end{table}
\addnew{The question of whether relativistic jets can be formed
 extracting the reducible energy of a rotating BH has been
 recurrently investigated in the last decades, both in the context of
 AGN jets \citep[e.g.][]{McKinney:2005,Hawley:2006,Komissarov2007a,
 Reynolds:2006,Garofalo:2009,Palenzuela:2011}, as well as in the context of gamma-ray bursts
 \citep[e.g.][]{Komissarov2009a,Nagataki:2009,Tchekhovskoy:2015,Nathanail:2016}. In
 this section, we employ the obtained solutions of the GSE in the
 split monopole magnetic field configuration to provide some analytic
 estimates of the total BZ power, as well as its distribution with
 latitude.}

We compute the total BZ power \citep[cf. also,][]{Thorne1986,
 Lee2000,Uzdensky2004,McKinney2004,Tanabe:2008,Tchekhovskoy2010, Penna2013} using
eq. (4.5) in \citet{Blandford1977}
\begin{align}
S^r\hspace{-2pt}(\theta)=\epsilon_0\omega\left(\Omega_{\mbox{\rm \text{\tiny BH}}}-\omega\right)\left(\frac{\Psi_{,\theta}}{r_+^2+a^2\cos^2\theta}\right)^2\left(r_+^2+a^2\right),
\label{eq:BZPower}
\end{align}
where $S^r$ is the radial energy flow. 
Then, employing
eq.\,(4.11) of \citet{Blandford1977}, we obtain the total power
outflow across the event horizon by integration of
eq.\,(\ref{eq:BZPower}):
\begin{align}
\begin{split}
P_{\rm tot}=&\int_0^{2\upi}\text{d}\phi\int_0^\upi\text{d}\theta\:S^r\hspace{-3pt}\left(\theta\right)\Sigma\sin\theta\\
=&\:\frac{\addnew{4}\upi\epsilon_0 a}{\Omega_{\mbox{\rm \text{\tiny BH}}}}\int_0^{\upi\addnew{/2}}\text{d}\theta\:\omega\left(\Omega_{\mbox{\rm \text{\tiny BH}}}-\omega\right)\frac{\sin\theta}{r_+^2+a^2\cos^2\theta}\:\left[\Psi_{,\theta}\right]^2,
\label{eq:Ptot}
\end{split}
\end{align}
where we have used the relations defined in eq. (\ref{eq:BHRotation}). The radial magnetic field component, $B^{\hat{r}}\equiv B_H$, in the local tetrad base of the ZAMO \citep[see, e.g.][]{Lee2000, Komissarov2009} may be defined as
\begin{align*}
\Psi_{,\theta}=\sqrt{g_{\theta\theta}}\sqrt{g_{\phi\phi}}\:B^{\hat{r}}=\sqrt{A}\sin\theta\:B^{\hat{r}}.
\end{align*}
At the event horizon ($\Delta=0$), this expression reduces to 
\begin{align}
\Psi_{,\theta}=\left(r_+^2+a^2\right)\sin\theta\:B^{\hat{r}}.
\label{eq:BrHor}
\end{align}
Following the ideas sketched in \citet{Lee2000} to obtain an
approximate expression for the BZ power, we assume that the ideal
fieldline angular velocity \citep{Blandford1977} is constant,
$\omega\simeq\OmegaBH/2$,\footnote{\addnew{This result was
 confirmed also numerically by \cite{Komissarov2001}; but see \cite{Pan:2015}.}} which in
combination with eq. (\ref{eq:BrHor}) and, $\epsilon_0 = 1/(4\upi)$
yields
\begin{align}
\begin{split}
P_{\rm tot}\simeq\:\frac{1}{\addnew{4}}a_*^2M^2&\left[1+\left(\frac{r_+}{a}\right)^2\right] \int_0^{\upi\addnew{/2}}\text{d}\theta\:\frac{\sin^3\theta}{\frac{r_+^2}{a^2}+\cos^2\theta}\:B_H^2,
\label{eq:PtotLee}
\end{split}
\end{align}
or, if we want to express the results in CGS units, we have
\begin{align}
\begin{split}
 P_{\rm tot}^{\rm cgs}\simeq \addnew{1.34\times 10^{51}}\text{erg\,s}^{-1} \times
 a_*^2&\,\left(\frac{M}{M_\odot}\right)^2\left(\frac{B_{\rm cgs}}{10^{15}\text{G}}\right)^2\times\\
\left[1+\left(\frac{r_+}{a}\right)^2\right]
\int_0^{\upi\addnew{/2}}&\text{d}\theta\:\frac{\sin^3\theta}{\frac{r_+^2}{a^2}+\cos^2\theta}\left(\frac{B_H}{B_{\rm cgs}}\right)^2,
\label{eq:UnitConversion}
\end{split}
\end{align}
where the magnetic field at the horizon is normalized to some reference value $B_{\rm cgs}$. In the following we will compare
different methods in order to evaluate the integral
\begin{align}
\tilde{P}\left(a\right)\equiv\:\int_0^{\upi\addnew{/2}}\text{d}\theta\:\frac{\sin^3\theta}{\frac{r_+^2}{a^2}+\cos^2\theta}\:B_H^2
\label{eq:PowIntegral}
\end{align}
in expression (\ref{eq:PtotLee}). In order to approximate $B_H$, we
proceed as follows. The obtained numerical results of the
split-monopole magnetosphere for different spin parameters,
$a_*\ge 0.7$ (see sec.\,\ref{sec:smconfigs}, especially
fig.\,\ref{fig:SMQuantities}) suggests that there exists a smooth
dependence of $\Psi(r_+,\theta)$ on the polar angle and $a$. For small
values of $a$, this is certainly the case
\citep{Blandford1977,MacDonald:1984MNRAS.211..313,McKinney2004}. We
have found the following fit function approximating the angular
dependency of the final and relaxed potential, as obtained from the
solution of the GSE along the inner radial boundary (outer event
horizon) for moderate to maximal values of $\addnew{a_*}$:
 \begin{align}
 \begin{split}
 \Psi_{\rm f}\left(r_+,\theta\right)\approx&\:\Psi_0\left(r_+,\theta\right)+c_1\times f\left(\addnew{a_*}\right)^{c_2}\times\sin\left(c_3\theta\right)\times\theta^{c_4}\\
 c_1=&\: 0.29,\qquad c_2=2.46, \qquad c_3=2.03,\qquad c_4=0.38, \\
 f\left(\addnew{a_*}\right)=&\:\frac{\addnew{a_*}}{1+\sqrt{1-\addnew{a_*}^2}}=\:\addnew{2}M\OmegaBH.
 \end{split}
 \label{eq:LCPotApprox}
 \end{align}
 The fit function reproduces the final values of $\Psi$
 computed with the GSE with an
 accuracy of
 $\left|\Psi_{\rm f}\left(r_+,\theta\right)-\Psi
 \left(r_+,\theta\right)\right|<0.02$. $B_H$ may be approximated by using the relation defined in eq. (\ref{eq:BrHor}):
 \begin{align}
 \begin{split}
 B_H\left(r_+,\theta\right)=\:\frac{1}{r_+^2+a^2}\frac{1}{\sin\theta}\frac{\partial\Psi_{\rm f}\left(r_+,\theta\right)}{\partial\theta}
 \label{eq:LCBApprox}
 \end{split}
 \end{align}
 For several values of $a$, we have
 integrated numerically eq.\,(\ref{eq:PtotLee}) using the numerical
 solution of the GSE for $B_H(r_+,\theta)$. 
 The results are plotted in
 fig.\,\ref{fig:Luminosity} (black crosses).

 Building upon \citet{Lee2000}, we employ eq.\,(\ref{eq:LCPotApprox}) to estimate $\langle\Psi_{,\theta}^2\rangle$ at
 the outer event horizon, finding
\begin{align}
\begin{split}
\langle\left[\Psi_{,\theta}\right]^2\rangle&=\int_0^{\frac{\upi}{2}}\text{d}\theta\sin\theta\left[\Psi_{,\theta}\right]^2\\
&=\frac{2}{3}-0.43 f\hspace{-2pt}\left(\addnew{a_*}\right)^{2.46}+0.18
f\hspace{-2pt}\left(\addnew{a_*}\right)^{4.9\addnew{2}}.
\end{split}
\label{eq:Psi_theta}
\end{align}
The integrand in eq. (\ref{eq:PowIntegral}) may then be approximated as follows:
\begin{align}
\begin{split}
\tilde{P}\left(a\right)\approx&\:\frac{\langle\left[\Psi_{,\theta}\right]^2\rangle}{\left[r_+^2+a^2\right]^2}\int_0^{\upi\addnew{/2}}\text{d}\theta\:\frac{\sin\theta}{\frac{r_+^2}{a^2}+\cos^2\theta}\\
=&\:\frac{\langle\left[\Psi_{,\theta}\right]^2\rangle}{\left[r_+^2+a^2\right]^2}\frac{a}{r_+}\arctan\frac{a}{r_+}
\end{split}
\label{eq:LeeIntegral}
\end{align}
\addnew{Inserting the latter expression into eq.\,(\ref{eq:PtotLee})
 we obtain a total power for the BZ process
\begin{align}
\begin{split}
P_{\rm tot}\simeq\:\frac{1}{4}
\frac{\langle\left[\Psi_{,\theta}\right]^2\rangle}{\left[r_+^2+a^2\right]}\frac{a}{r_+}\arctan\frac{a}{r_+}\:,
\end{split}
\label{eq:LeePower1}
\end{align}
or, equivalently,
\begin{align}
\begin{split}
P_{\rm tot}\simeq\:\frac{\epsilon_0\upi}{2}\langle\left[\Psi_{,\theta}\right]^2\rangle
\OmegaBH M (1+4M^2\OmegaBH^2)\arctan{\left(2\OmegaBH M\right)}\:,
\end{split}
\label{eq:LeePower2}
\end{align}
which, expanding in series of $\OmegaBH$ and retaining terms up
to second order yields
\begin{align}
\begin{split}
P_{\rm tot,2}\simeq\:\epsilon_0 \frac{2\upi}{3}M^2\OmegaBH^2.
\end{split}
\label{eq:LeePower2exp}
\end{align}

In their study of the spin dependency of the power of the BZ
 process, \citet{Tchekhovskoy2010} introduce expansions
 of the cumulative power (i.e., the angular power density integrated
 up to a certain angle $\theta_{\rm j}$ instead of up to $\upi/2$ as in
 eq.\,\ref{eq:Ptot}) in terms of $\OmegaBH$. The resulting
 second, 4th, and 6th order accurate expressions of the total power
 (i.e. the cumulative power up to $\theta_{\rm j}=\upi/2$) are in our
 notation and units\footnote{\addnew{\cite{Tchekhovskoy2010} employ
 units in which $\epsilon_0=1$.}} 
\begin{align}
P_{\rm tot,BZ2}&=\epsilon_0 \frac{2\upi}{3}M^2\OmegaBH^2\Psi_{\rm tot}^2\label{eq:BZ2}\:,\\
P_{\rm tot,BZ4}&=\epsilon_0 \frac{2\upi}{3}\left[M^2\OmegaBH^2+b_1 M^4\OmegaBH^4\right]\Psi_{\rm tot}^2\label{eq:BZ4}\:,\\
P_{\rm tot,BZ6}&=\epsilon_0 \frac{2\upi}{3}\left[M^2\OmegaBH^2+b_1 M^4\OmegaBH^4+b_2 M^6\OmegaBH^6\right]\Psi_{\rm tot}^2\:,\label{eq:BZ6}
\end{align}
where $\Psi_{\rm tot}$ corresponds to the total flux between
$\theta=0$ and $\theta=\upi/2$, i.e. $\Psi_{\rm tot}=1.0$. We note that
the second-order accurate expression of this work
(eq.\,\ref{eq:LeePower2exp}) and of \cite{Tchekhovskoy2010},
eq.\,(\ref{eq:BZ2}), are identical.
The coefficient of the term proportional to $\OmegaBH^4$,
$b_1=8(67-6\upi^2)/45\simeq 1.38$, is computed analytically. For the
6th order accurate expression in eq. (\ref{eq:BZ6}),
\cite{Tchekhovskoy2010} obtains $b_2=-9.2$ from a least-squares fit to
their full analytic formulae\footnote{\addnew{We note that
 \cite{Pan:2015} obtain the same value of $b_1$ as
 \cite{Tchekhovskoy2010}, but $b_2=-11.09$.}}.
As an alternative to the coefficients employed in
\cite{Tchekhovskoy2010}, we may compute the coefficients employed in
eq. (\ref{eq:BZ6}) according to our specific numerical solution
derived with the chosen fit function (eq. \ref{eq:LCPotApprox}). The
resulting coefficients are then $b_1=0.81$ and
$b_2=-5.62$. For brevity, we refer to
this parameter set as BZ6b hereafter. We shall point out that the
expression (eq.\,\ref{eq:BZ6}) is, indeed, not formally sixth-order
accurate. It neglects the (typically small) corrections introduced by
approximating the fieldline angular velocity as $\omega\simeq
\OmegaBH/2$. An expansion of $\omega$ accurate up to $O(\OmegaBH^6)$
can be found in \cite{Pan:2015}. 

We compare the approximations obtained by \cite{Tchekhovskoy2010} to
ours in fig.\,\ref{fig:Luminosity}. We find that the approximation as
suggested by \citet{Lee2000} is equally good or comparable to the
suggested 6th order approximation (eq.\,\ref{eq:BZ6}) up to spin
parameters of $a_*\leq 0.98$, and more accurate than the second and
4th order formulae (eqs.\,\ref{eq:BZ2} and \ref{eq:BZ4},
respectively) for the entire range of $a_*$. For extremal spins above
this threshold, eq. (\ref{eq:BZ6}) yields very accurate
estimates. However, using our fit parameters in the 6th order
approximation of eq.\,(\ref{eq:BZ6}), we obtain even more accurate
results as compared to the remaining numerical models.}

\addnew{The global accuracy of
the results is assessed in tab.\,\ref{tab:rms}. The table shows the root
mean square (RMS) deviations of the different approximations employed
to compute the total BZ power, i.e.
\begin{equation}
E=\left(\sum_{i} (P_{{\rm GSE},i} - P_i)^2\right)^{1/2},
\end{equation}
where $P_{{\rm GSE},i}$ and $P_i$ represent, respectively, the power
computed numerically from the solution of the GSE and the estimation
of the total BZ power obtained with eq.\,(\ref{eq:PtotLee}),
eq.\,(\ref{eq:BZ2}), eq.\,(\ref{eq:BZ4}) or eq.\,(\ref{eq:BZ6}) using
the original parameters of \cite{Tchekhovskoy2010} or our own
parameters (BZ6b). In order to facilitate the comparison, all the RMS
errors are normalized to the RMS deviations of the model BZ6b. The
relatively simple approximation of eq.\,(\ref{eq:PtotLee}) displays a
RMS error which is $\lesssim 30\%$ larger than the original 6th order
approximation to estimate the BZ total power (eq.\,\ref{eq:BZ6}). We
observe that the 4th order approximation (eq.\,\ref{eq:BZ4}) deviates
more from the data than even the second-order estimate
(eq.\,\ref{eq:BZ2}) or the BZ total power estimation using our
eq.\,(\ref{eq:PtotLee}). This is not surprising, since it was also
anticipated in \cite{Tchekhovskoy2010}.}
\begin{figure}
	\centering
	\includegraphics[width=0.49\textwidth]{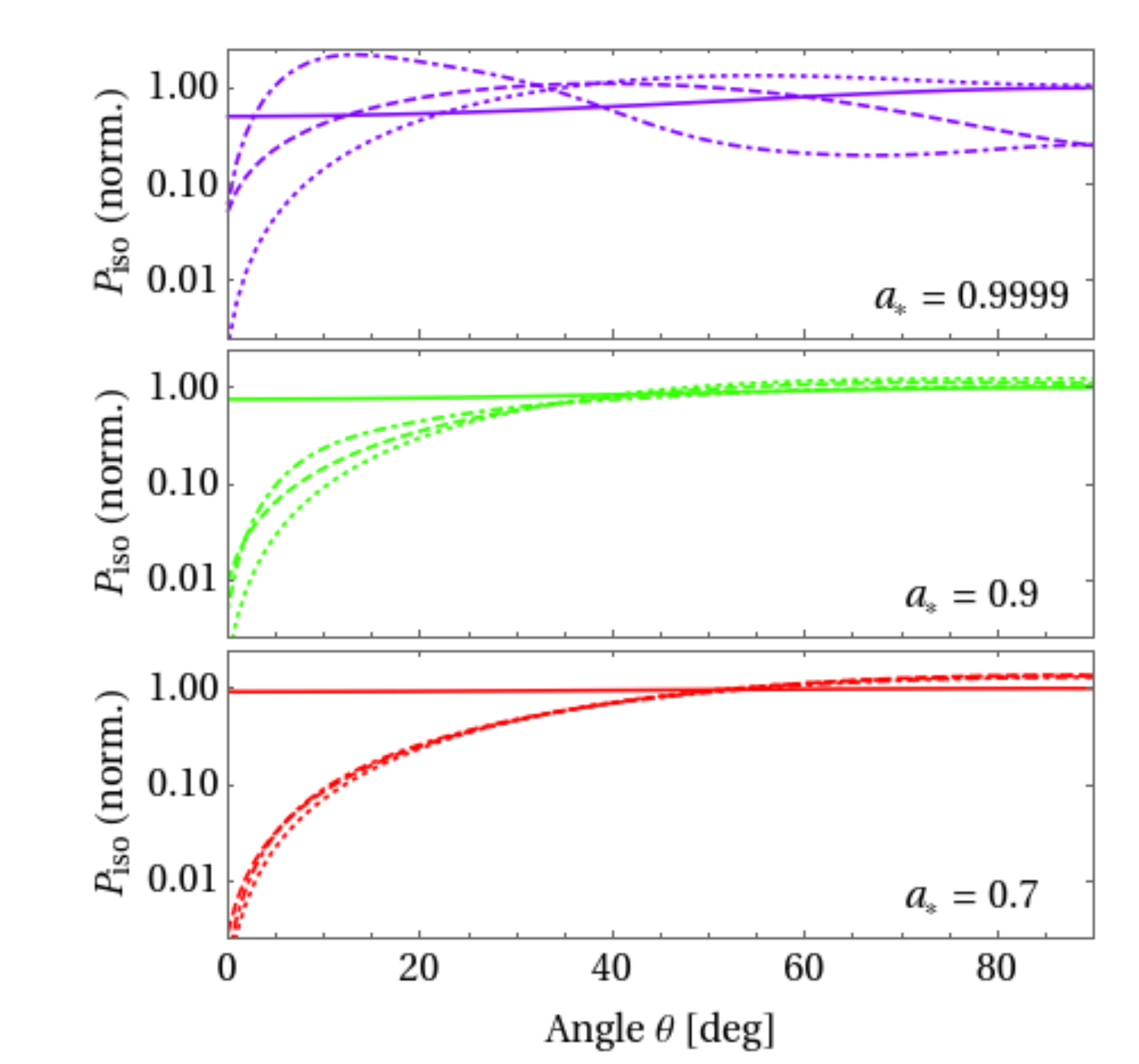}
	\caption{Isotropic power (eq. \ref{eq:PIso}) for different
 spin parameters $a_*=\left\{0.7,0.9,0.9999\right\}$. The
 \addnew{solid} lines refer to the \addnew{approximated
 analytic} integration of eq.\,(\ref{eq:PIsoanal}) plotted
 for intervals of
 $\left|\theta^+-\theta^-\right|=\upi/200$. The
 \addnew{dashed} lines show the direct integration of
 eq. (\ref{eq:PIso}) using the midpoint rule
 \addnew{(i.e. using eq.\,\ref{eq:AngularApprox}). The dotted
 and dot-dashed lines represent the angular power
 derivation following the 4th, and 6th order approximations
 \addnew{given by eqs.\, (\ref{eq:PisoBZ4}) and
 (\ref{eq:PisoBZ6}), respectively}. All lines are
 normalized to the maximum value of the isotropic power
 provided by eq.\,(\ref{eq:PIsoanal}) in the interval
 $[0,\upi/2]$.} }
	\label{fig:Luminosity2}
\end{figure}

If the power of the BZ process drives a collimated relativistic
outflow from the neighborhood of the BH, a distant observer may only
see an small angular patch of the whole outflow due to relativistic
beaming. Thus, \addnew{following a common practice for the approximate
  estimation of the angular energy distribution in GRB ejecta
  \citep[e.g.][]{Janka2006,Mizuta2009,Lazzati2009} that avoids
  performing a complete (and much more involved) radiation transport
  problem \citep{Broderick2003,Miller2003,Birkl2007,Cuesta-Martinez2015},} it is useful to define an
equivalent isotropic power $P_{\rm iso}\addnew{(\theta)}$ in each
(narrow) angular region $\Delta \theta = \theta^+ - \theta^-$
\addnew{centered around $\theta$ (i.e.
  $\theta=(\theta^+ - \theta^-)/2$)} as:
\begin{align}
\begin{split}
P_{\rm
 iso}\left(\theta\right)=&\:\frac{4\upi\epsilon_0}{\cos\theta^--\cos\theta^+}\frac{a}{\OmegaBH} \times\\
&\int_{\theta^-}^{\theta^+}\text{d}\theta\:\omega\left(\OmegaBH-\omega\right)\frac{\sin\theta}{r_+^2+a^2\cos^2\theta}\:\left[\Psi_{,\theta}\right]^2
\end{split}
\label{eq:PIso}
\end{align}
This integral may be further simplified by using
eq. (\ref{eq:Psi_theta}) to replace $\left[\Psi_{,\theta}\right]^2$ by
its angular average. The remaining integral is the same as in
eq. (\ref{eq:LeeIntegral}) and can be solved
analytically \addnew{yielding,
\begin{align}
\begin{split}
P_{\rm iso}^{(1)}\left(\theta\right)=&\:\frac{2\upi\epsilon_0}{\cos\theta^--\cos\theta^+}\frac{\OmegaBH}{r_+^2+a^2} \left[\Psi_{,\theta}\right]^2\times\\
&\left\{\arctan{\left(\frac{a}{r_+}\cos\theta^-\right)} - \arctan{\left(\frac{a}{r_+}\cos\theta^+\right)} \right\}.
\end{split}
\label{eq:PIsoanal}
\end{align}
} Alternatively, eq. (\ref{eq:PIso}) can be computed using the
midpoint approximation and plugging in for $\Psi_{,\theta}$ the
angular derivative of $\Psi_{\rm f}\left(r_+,\theta\right)$ as given
in eq. (\ref{eq:LCPotApprox}). The angular dependence of the isotropic
power then reduces to
\begin{align}
P_{\rm iso}^{(2)}\left(\theta\right) =\addnew{\frac{\upi\epsilon_0\Delta\theta}{\cos\theta^+-\cos\theta^-}} a \OmegaBH\frac{\sin\theta}{r_+^2+a^2\cos^2\theta}\:\left[\Psi_{,\theta}\right]^2.
\label{eq:AngularApprox}
\end{align}
\addnew{Figure} \ref{fig:Luminosity2} shows the isotropic power
employing eq.\,(\ref{eq:PIsoanal}) plotted for intervals of
$\Delta\theta=\left|\theta^+-\theta^-\right|=\upi/200$, as well as the
angular approximation given in eq. (\ref{eq:AngularApprox}). 
\addnew{For comparison, we also plot the isotropic equivalent power
 distribution following the 4th and 6th order approximations of
\citet{Tchekhovskoy2010}. Precisely, we show 
\begin{align}
P_{\rm iso,BZ4}(\theta)=&\frac{2}{\cos\theta^+-\cos\theta^-}
\left(P^{\rm cum}_{\rm BZ4}(\theta^+) - P^{\rm cum}_{\rm BZ4}(\theta^-)\right)\:,\label{eq:PisoBZ4}
\end{align}
and
\begin{align}
P_{\rm iso,BZ6}(\theta)=&\frac{2\Delta\theta}{\cos\theta^+-\cos\theta^-}
\frac{dP_{\rm BZ6}}{d\theta}(\theta)\:, \label{eq:PisoBZ6}
\end{align}
where the 4th order accurate expression of the cumulative power as a
function of the angle is \citep[cf.][eq.\,B6]{Tchekhovskoy2010}
\begin{eqnarray}
P^{\rm cum}_{\rm BZ4}(\theta )=\hspace{-0.3cm}&\displaystyle\frac{1}{270} \upi \Omega_{\rm \text{\tiny BH}}^4 
\Bigg(\hspace{-0.3cm}&90 \left(3 \upi^2-32\right) \cos{(\theta)}+\notag\\
&&5\left(194-21 \upi^2\right) \cos{(3\theta )}+\notag\\
&&9 \left(3 \upi^2-26\right) \cos (5 \theta )+\\
&&32 \left(67-6 \upi^2\right)\Bigg)+\notag \\
\hspace{-0.25cm}&\displaystyle\frac{1}{3} \upi \Omega_{\rm \text{\tiny BH}}^2 
\Bigg(\hspace{-0.3cm}&\hspace{-0.1cm} 4\sin^4\left(\frac{\theta}{2}\right) (\cos{(\theta )+2)}\Bigg)\:,\notag
\end{eqnarray}
and the differential power per unit angle is given by 
\begin{align*}
\frac{dP_{\rm BZ6}}{d\theta}(\theta) =&\upi a \Omega_{\mbox{\rm \text{\tiny BH}}}\frac{\sin\theta}{r_+^2+a^2\cos^2\theta}\:\left[\Psi^{\rm (6)}_{,\theta}\right]^2,
\end{align*}
with $\Psi^{\rm (6)}_{,\theta}$ the angular derivative of the 6th order
accurate approximation for the magnetic flux \citep[cf.][eqs.\,C1 and C2]{Tchekhovskoy2010}
\begin{align*}
\Psi^{\rm (6)} = &\: \Psi_0(r_+,\theta) + 
16\Omega_{\mbox{\rm \text{\tiny BH}}}^2\frac{8(67 - 6\upi^2 )}{45}
 \sin^2\theta \cos\theta+\\ 
&\: \Omega_{\mbox{\rm \text{\tiny BH}}}^4\sin^2\theta \: \left(26.12\cos^{25} \theta+22.72\cos^{7}\theta
+\right. \\
& \qquad\qquad\:\:\:\: \left.13.54 \cos^3\theta + 2.08 \cos\theta\right).
\end{align*}
Looking at the central and bottom panels of
fig.\,\ref{fig:Luminosity2}, it is evident that estimating the
isotropic equivalent power with the approximation leading to
eq.\,(\ref{eq:PIsoanal}) is not optimal, especially for
$\theta\lesssim 40^\circ$ and moderate to large values of $a_*$. This
is not surprising, since the average performed to compute (eq.\,\ref{eq:Psi_theta}) extends over
the whole interval $\theta\in[0,\upi/2]$, while the isotropic power is
evaluated for a relatively small angular patch, with angular extension
$\Delta\theta$. Our isotropic power estimate employing the mid-point
rule (eq.\,\ref{eq:AngularApprox}) yields much better results. It is
closer to the 6th-order accurate estimation of eq.\,(\ref{eq:PisoBZ6})
than the 4th-order accurate estimation of
eq.\,(\ref{eq:PisoBZ4}). However, it falls short to predict the
isotropic power for $\theta\lesssim 25^\circ$ and almost maximal
values of the BH spin ($a_*=0.9999$). At small or moderate values of
$a_*\lesssim 0.7$, the isotropic power estimations yield
quantitatively similar results, regardless of the approximation
employed to compute the angular distribution of the BZ power (with the
notable exception of $P_{\rm iso}^{(1)}(\theta)$). The differences
show up more clearly as the value of $a_*$ grows. There is, however, a
consistent trend in all cases:} larger values of the BH spin parameter
$a_*$ show larger powers closer to the axis of rotation. Indeed, we
observe a transition in the curves of $P_{\rm iso}$. For
\addnew{$a_*> 0.9$} the maximum value of $P_{\rm iso}$ shifts from
$\theta=90^\circ$ to lower latitudes. For \addnew{nearly} maximally rotating BHs, the
maximum isotropic equivalent power happens for
$\theta\simeq 10^\circ$. Regardless of the exact location of the
maximum, fig.\,\ref{fig:Luminosity2} clearly shows that a distant
observer would not see a maximum power for events seen
``head-on''. Since the value of $P_{\rm iso}$ grows very steeply from
zero, for events generated out of BHs with $a_*\gtrsim 0.9$, it is
much more likely to observe luminous events when observing them at
angles $\theta\gtrsim \addnew{10^\circ}$.

\section{Discussion}
\label{sec:discussion} 
Motivated by the study of relativistic outflows from spinning black
holes \citep{Blandford1977}, magnetospheric force-free electrodynamics
for static and axisymmetric spacetimes have become a matter of active
research \citep[e.g.][]{Camenzind2007,
 Beskin2010}. \cite{appl:1993A&A...274..699} found the first
non-linear analytical solution for a cylindrically collimated,
asymptotic flux distribution, in special relativity. Until now, most
of these studies have been done in the context of \textit{open field}
configurations, primarily because of its relevance to the jet
problem. However, complex magnetic field topologies encompassing
closed fieldlines may also develop in the course of the dynamical
evolution arising from the accreting BHs
\citep{Goodman2008,Parfrey2016}. Indeed, it has been
encountered in simulations of neutron star mergers
\citep[e.g.][]{Rezzolla2011,Kiuchi2014}, that the post merger BH
magnetic fields are not necessarily of split-monopole or paraboloidal
type. The exact topology of the magnetic field in the BH magnetosphere
is of paramount importance to set the efficiency of energy extraction
from the central compact object. This extraction of energy is supposed
to be channeled out along the low-density funnel developed in the
course of the merger around the rotational axis of the system.
Especially in the low density funnel, field reversals have been
encountered, possibly limiting the efficiency of outflow production.

Studying these phenomena requires accurate initial data for
magnetospheric configurations and motivates us to build both
transparent and versatile initial data solvers. In particular, it
requires a proper characterization of the numerical methodology to
solve the GSE. 

The numerical method we propose splits into three basic blocks: (1)
The finite difference solution of the GSE in each of the subdomains
set by the light surfaces in the magnetosphere, (2) the matching of
the solutions across the light surfaces to obtain regular functions,
and (3) the build-up or update of the functional tables for
$\omega(\Psi)$ and $II'(\Psi)$. 

We have shown that the convergence of the presented numerical
technique greatly depends on a suitable selection of finite difference
discretization around the LS and, hence, the diagonal dominance of the
coefficient matrix of the SOR solver. \addnew{Numerical artifacts
 develop around the LS if the 'smoothing-across-subdomains'
 techniques are used \cite[as in
 e.g.][]{Nathanail2014,Pan:2017}. These artifacts slow down the
 convergence of the GSE solution significantly, and the quality of
 the results is limited} (see
fig.\,\ref{fig:ConvergenceL}). \addnew{However, } we find the choice
of an inward/outward biased second order discretization around and
with respect to the LS to be the most efficient setup to achieve fast
convergence with convenient overrelaxation and no need for additional
smoothing of the potential $\Psi$ across the LS.

We have extended the strategy of employing eq. (\ref{eq:GSReduced}) to
relax the current $II'\left(\Psi\right)$ \citep{Uzdensky2004} also to
the potential function $\omega$. This allows us to use the error of
the GSE at the LS (eq.\,\ref{eq:GSLightCylinder}) as an additional
measure of convergence for the presented numerical method. Despite the
\addnew{fact that we can also incorporate a local mesh refinement
  around the ILS}, the method is, however, limited by the ability to
fit sufficient grid cells between the BH horizon and the ILS. In the
presented tests, we were able to ensure sufficient cells around the
ILS for values as low as $a_*=0.5$. Nevertheless, the numerical
solution of the GSE for spin factors lower than $a_*=0.7$ is perhaps
unnecessary. As shown in fig.\,\ref{fig:SMQuantities}, \addnew{even
  for a BH spin as large as} $a_*=0.7$, the overall solution
approached the initially guessed potential functions, which are the
(exact) solutions of the case $a_*=0$. \addnew{Thus, we find that with
  our method it is possible to explore the region very close to the
  horizon for rotating BHs with $a_*\gtrsim 0.7$. Resolving this
  region, which is near the location of the zero space-charge, and
  which is expected to be the source of the pairs that will populate
  the BH magnetosphere \citep{Globus:2014}, is very important for
  models of jet formation \citep[as it is in pulsar magnetospheres;
  e.g.][]{Belyaev2016}.}

\addnew{In the literature, the solutions found by the numerical
 relaxation of the GSE seem to evolve towards a unique solution as
 long as the boundary conditions, especially the thin-disk
 assumption, hold \citep{Contopoulos2013,Nathanail2014}. We have
 observed and discussed (see sec. \ref{sec:UpdatesPotentials}) the
 relaxation of both potential functions $\omega$ and $I$ as a
 necessity for convergence under the criterion given in appendix
 (\ref{sec:convergence}). The relaxation of only one of these two
 functions may converge under the residual function
 $\mathcal{R}_\Psi$, but show non-convergence under the
 (mathematical) measure $\mathcal{R}_{LC}$, employing the GSE at the
 location of its singular surfaces (eq. \ref{eq:RLC}). Hence, we
 cannot disprove the uniqueness of the solutions reproduced in
 sec. \ref{sec:NumericalResults}.}
 
 \addnew{Proving the uniqueness of the solution of the GSE, given a set of boundary conditions,
 is not an easy task. The customary way of demonstrating uniqueness is to find
 a maximum principle. If no maximum principle can be found multiple solutions may arise for a
 given set of boundary conditions. Such an example can be found in \cite{Akgun:2018}, where
 the authors solved the Newtonian GSE equations without rotation for the case of a neutron star 
 with a twisted magnetosphere, in some cases finding numerically multiple solutions with identical boundary conditions. 
 In some particular cases uniqueness can be proven as in the case of current-free configurations \citep[see e.g.][]{Akgun:2018}
 or for small twists \citep{bineau1972}. Regarding black hole magnetospheres, \citet{Pan:2017} have investigated the uniqueness of the
 solution of the GSE. These authors found that if the field lines cross smoothly the LSs (which is a 'constraint
 condition' the solutions must satisfy), the boundary conditions at
 the horizon and at infinity are not independent. Therefore, for a given pair of functions $\omega(\Psi)$ and $I (\Psi)$
 the boundary conditions are uniquely defined. Although this does not completely prove the uniqueness of
 the solution for given boundary conditions, it is a significant step in this direction. However, for the asymptotically
 uniform field, there is a variety of possibilities regarding
 uniqueness. Time-dependent simulations
 \citep[e.g.][]{Komissarov:2005,Komissarov2007a,Yang:2015} apparently
 converge to a unique solution. Several analytic studies find a
 unique perturbative solution that agrees with GRMHD simulations
 \citep{Beskin:2013,Pan:2015,Gralla:2016}.
}

As a byproduct of the study of split-monopole magnetospheres, we have
provided an approximation for the potential $\Psi$ at the outer event
horizon for different values of $a$ (eq.\,\ref{eq:LCPotApprox}). In
section \ref{sec:BZpower}, we examined the angular resolution as well
as the total value of the power outflow
\citep[cf.][]{Lee2000,Uzdensky2004} employing
eq. (\ref{eq:LCPotApprox}). Especially for higher values of the BH
spin parameter $a$ one finds a higher total power of the BZ process
with more isotropic power provided closer to the axis of rotation and,
hence, in the regions which are presumably critical for the production
of BZ jets. \addnew{Our estimations of the power of the BZ process
 using the fit formula (eq.\,\ref{eq:PtotLee}) deviate less than
 1\% from the \emph{exact} power computed numerically from the
 solutions of the GSE equations for the potential $\Psi$. Remarkably,
 this compares fairly well with the 4th order accurate expression
 of \cite{Tchekhovskoy2010} (their eq. B6), since their formula requires more
 than a factor of 3 correction to reproduce their numerical results
 for high BH spin (namely, $a_*\gtrsim 0.95$), as the authors point
 out. Thus, our relatively simple estimate of the BZ power provides
 estimates quantitatively comparable to the 6th order accurate
 expression of \citeauthor{Tchekhovskoy2010}
 (\citeyear{Tchekhovskoy2010}; their eq.\,9).}

\addnew{We also find that using our fit formula
 (eq.\,\ref{eq:LCPotApprox}) to compute the angular derivative of the
 flux (which is proportional to the radial component of the magnetic
 field evaluated at the BH horizon; eq.\,\ref{eq:LCBApprox}) is an
 excellent approach to estimate the isotropic equivalent power as a
 function of the latitude. Our simple estimate is competitive with
 the 6th order accurate estimation of \cite{Tchekhovskoy2010} for BH
 spins $a_*\lesssim 0.98$. However, for nearly maximally rotating BH
 ($a_*\gtrsim 0.9999$), our isotropic equivalent power estimate falls
 short by factors $2-3$ with respect to the estimation employing a
 6th order accurate formula. Remarkably, even for such large values
 of $a_*$, our mid-point approximation for the isotropic BZ
 luminosity (eq.\,\ref{eq:AngularApprox}) is better than the 4th
 order accurate estimation of eq.\,(\ref{eq:PisoBZ4}).}

\addnew{The stability of most of the stationary solutions found in
 this paper (and in the preceding literature in the field) has been
 assessed by means of time-dependent FFDE or GRMHD simulations
 \citep[e.g.][]{Komissarov2002,Komissarov2004,Tchekhovskoy2010} with a fixed background metric (the
 one provided by the BH). However, the stability of the solutions in
 cases where the space time may evolve due to the feedback between
 the BH and its magnetosphere has not been assessed so far.} One
application of the presented solving scheme will be the use of these
configurations as initial data for time evolution simulations of
\textit{dynamical} spacetimes on the \textit{Carpet} grid of the
\textit{Einstein Toolkit}. The discussed test cases are thought to be
especially applicable in combination with recent methods to support
rapidly rotating Kerr BHs in numerical simulations \citep{Liu2009} and
their numerically stable magnetohydrodynamic evolution
\citep{Faber2007}. \addnew{The current practice of evolving spacetimes
 without excising the BHs \citep[for GRMHD simulations,
 cf.][]{Faber2007} requires highly accurate initial data, especially
 at the BH apparent event horizons}. The behavior under time
evolution \addnew{of \textit{dynamical} spacetimes (without excising
 BHs) with the respective time-dependent feedback on the
 electromagnetic force-free fields} will be an indicator on the
stability of the found solutions of the GSE and may foster further
optimization of the proposed numerical procedure. The results of the
application of this methodology will be the subject of our subsequent
work.

\section{Acknowledgements}
\addnew{We thank the referee, Prof. I. Contopoulos, for his
 constructive comments and criticism.} We kindly acknowledge Amir
Levinson for his careful reading and feedback on the draft of this
paper. We acknowledge the support from the \textit{European Research
 Council} (grant CAMAP-259276) and the partial support of grants
AYA2015-66899-C2-1-P and PROMETEO-II-2014-069. JM acknowledges the
\textit{Grisolia} Grant GRISOLIAP/2016/097 and a Ph.D. grant of the
\textit{Studienstiftung des Deutschen Volkes}. PC acknowledges the
Ramon y Cajal program (RYC-2015-19074) supporting his research.

\bibliographystyle{mnras}
\bibliography{Literature}

\appendix
\section{Technical notes}
\label{sec:technical_notes} 

\subsection{Numerical convergence criteria}
\label{sec:convergence}
%
\begin{figure}
	\centering
	\includegraphics[width=0.42\textwidth]{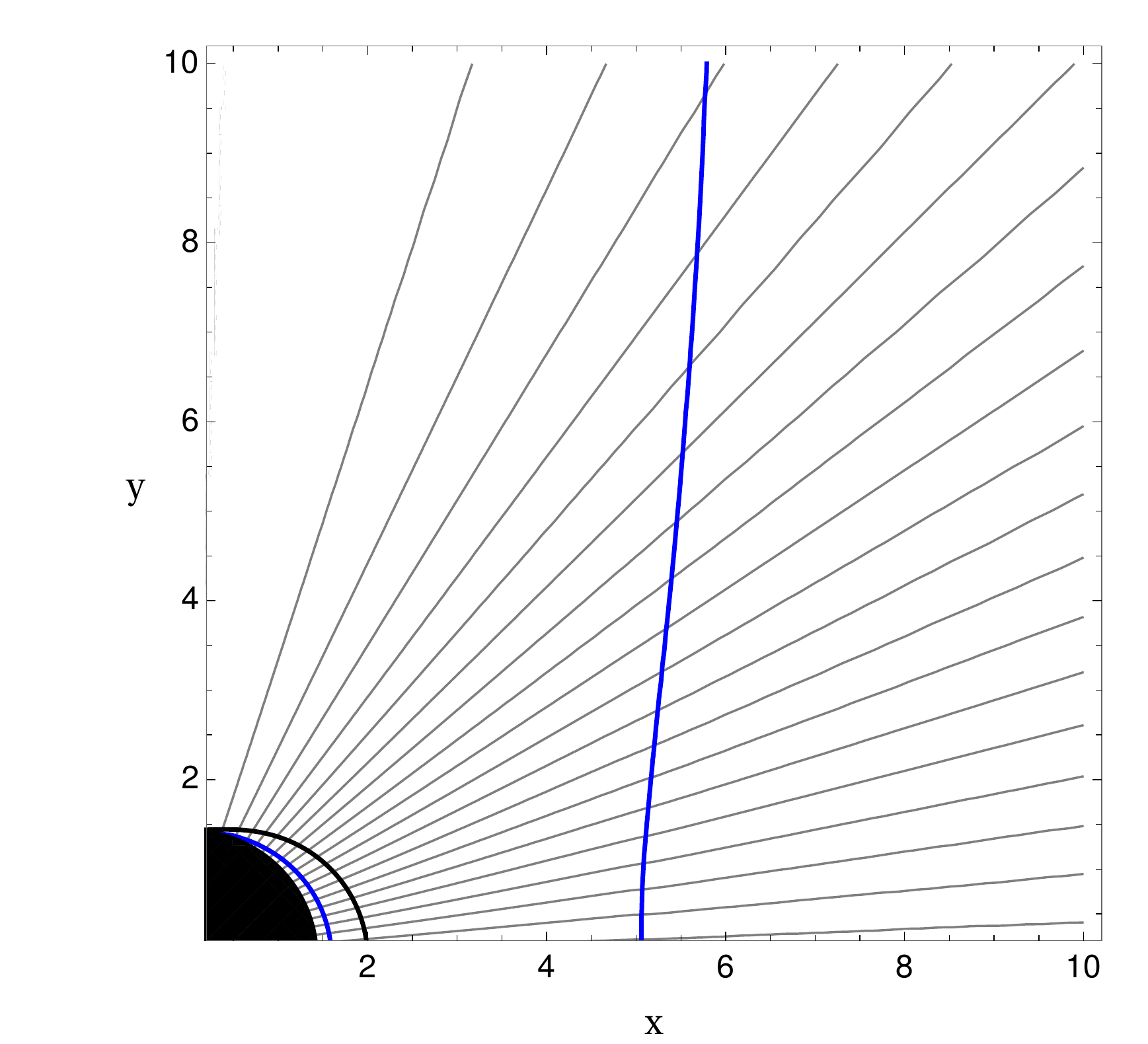}
	\includegraphics[width=0.42\textwidth]{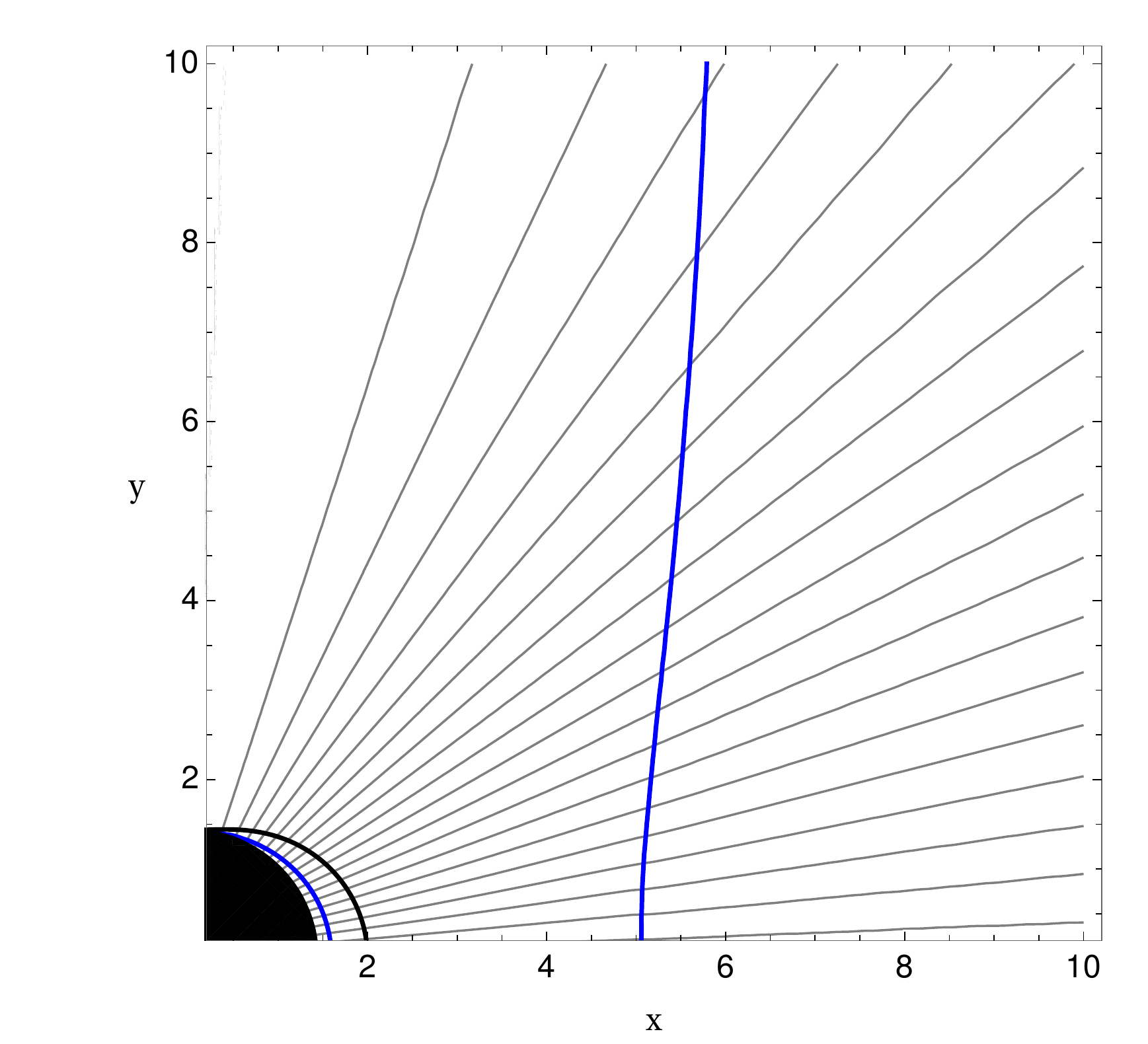}
	\includegraphics[width=0.42\textwidth]{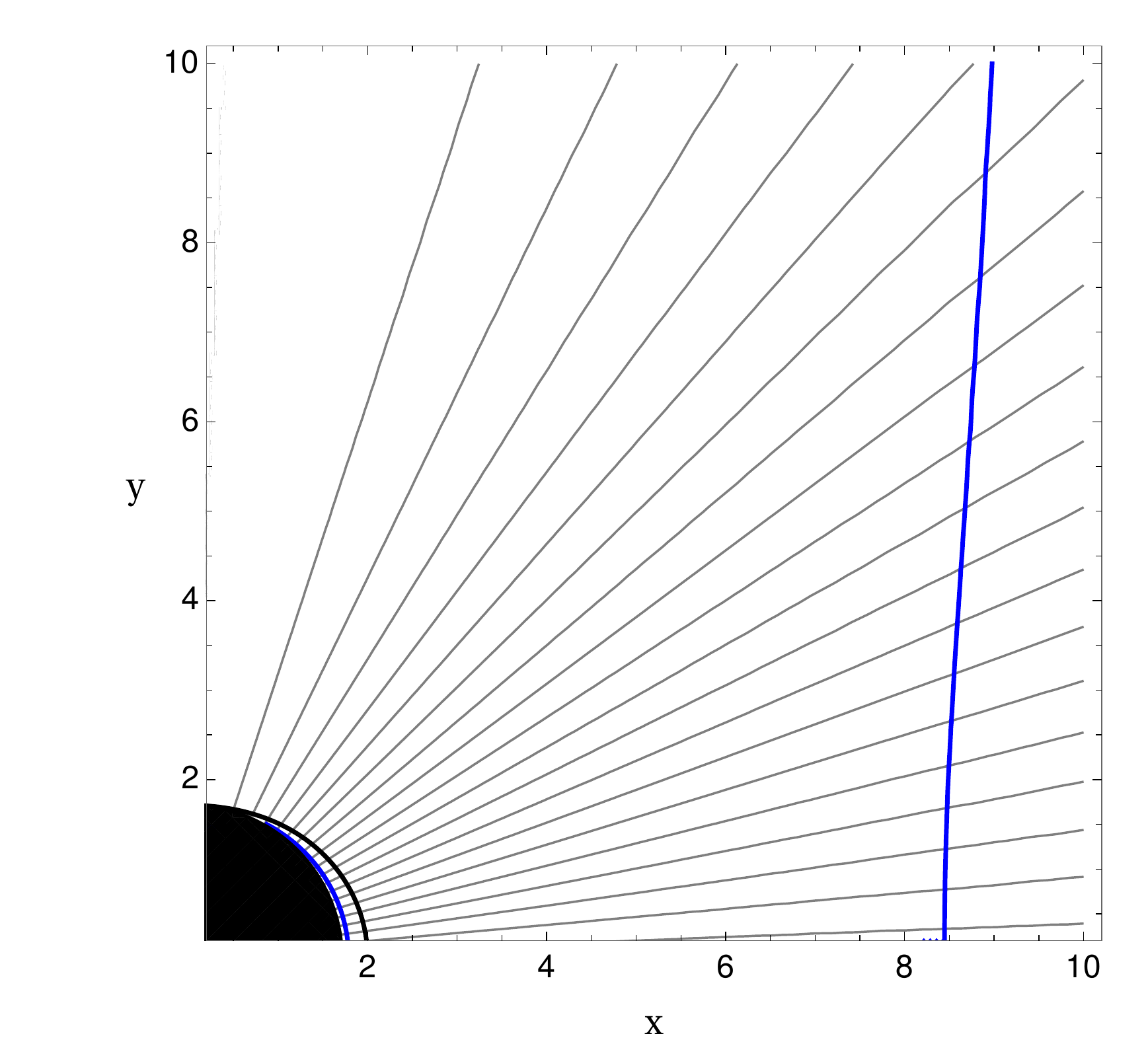}
	\caption{Distribution of the magnetic flux $\Psi$ in the
 vicinity of a black hole rotating at $a_*=0.999$
 (top), $a_*=0.9$ (middle) and $a_*=0.7$
 (bottom) after converging under the criteria in
 eq. (\ref{eq:ConvCriterion}) on the physical domain
 $[r_+,\infty] \times [0,90^\circ]$, covered with a numerical
 grid
 $\left[n_r\times n_\theta\right]=\left[200\times
 64\right]$. Both functions $\omega\left(\Psi\right)$ and
 $II'\left(\Psi\right)$ are relaxed throughout the iterative
 procedure. The location of the ergosphere is represented by
 the black line, the two LS are drawn as blue lines.}
	\label{fig:SMConfigs}
\end{figure}
As shown in figs.\,\ref{fig:ConvergenceL} and
\ref{fig:ConvergenceLLong}, the $L^\infty$ norm of the residual of the
SOR scheme decreases rapidly for the applied biased stencil at the
light surfaces. However, we find that a smooth solution for $\Psi$ in
the entire domain (cf. sec.\,\ref{sec:ligthcylinders}) requires an
additional constraint imposed by the error $\mathcal{R}_{LC}$ at the
LS. \addnew{Kinks across a LS may remain}
present even though the solution converged under
$\mathcal{R}_\Psi$. Furthermore, especially for non-extremal spin
parameters $a$ we observe that both $\mathcal{R}_{LC}$ and
$\mathcal{R}_{\Psi}$ are very small already in the first iterations of
the solving routine. An exclusive focus on the residual
$\mathcal{R}_{\Psi}$ may, hence, rapidly trigger a convergence
decision without a full relaxation of $\omega$ and $II'$. In these
cases, kinks remain across the LS. For the shown tests we demand the
simultaneous fulfillment of the following conditions as a convergence
criterion:
\begin{align}
\mathcal{R}_{\Psi} < 10^{-6} \qquad\wedge\qquad \mathcal{R}_{LC} < 5 \times 10^{-4/\sqrt{a}}
\label{eq:ConvCriterion}
\end{align}
Figure (\ref{fig:SMConfigs}) shows selected split-monopole
configurations after condition \ref{eq:ConvCriterion} has been
reached. For different values of the black hole spin $a$, the required
iterations to reach convergence are shown in table \ref{tbl:Convergence}.

\begin{table}
	\centering
\begin{tabular}{c|c}
		Spin ($a_*$) & Iterations until convergence\\
		\hline
		0.9999	& 50.161 \\
		0.999	& 29.100 \\
		0.99	& 64.117 \\ 
		0.9	& 43.153 \\
		0.8	& 12.100 \\
		0.7	& 19.129 \\
		\hline
		\hline
	\end{tabular}
	\caption{Number of iterations until the convergence criterion (\ref{eq:ConvCriterion}) is reached for the \textit{split-monopole} setup described in section \ref{sec:smconfigs}.}
	\label{tbl:Convergence}
\end{table}

\label{lastpage}

\end{document}